\PassOptionsToPackage{table}{xcolor}
\documentclass[manuscript,screen]{acmart}

\usepackage{amsmath,amsfonts}
\usepackage{algorithmic}
\usepackage{graphicx}
\usepackage{textcomp}
\usepackage{xcolor}
\usepackage{xspace}
\usepackage{fontawesome}
\usepackage{multirow}
\usepackage{rotating}
\usepackage{tikz}
\usepackage{soul}
\usepackage{makecell}
\usepackage{url}
\usepackage{enumitem}
\usepackage{booktabs}
\usepackage{fontawesome}
\usepackage{pifont}
\usepackage{tcolorbox}
\usepackage{tabularx}
\usepackage[table]{xcolor} 
\usepackage[normalem]{ulem} 
\usepackage[caption=false]{subfig}
\usepackage{listings}
\usepackage{afterpage}

\newcommand{\reducedstrut}{\vrule width 0pt height \ht\strutbox depth .9\dp\strutbox\relax}
\newcommand{\pink}[1]{%
  \begingroup
  \setlength{\fboxsep}{0pt}%
  \colorbox{red!20}{\reducedstrut#1\/}%
  \endgroup
}

\newcommand{\ie}{\emph{i.e.,}\xspace}
\newcommand{\eg}{\emph{e.g.,}\xspace}
\newcommand{\etc}{etc.\xspace}
\newcommand{\etal}{\emph{et~al.}\xspace}
\newcommand{\secref}[1]{Section~\ref{#1}\xspace}

\newcommand{\figref}[1]{Fig.~\ref{#1}\xspace}

\newcommand{\tabref}[1]{Table~\ref{#1}\xspace}

\newcommand{\ds}{\emph{developer}-\emph{specific}\xspace}
\newcommand{\os}{\emph{organization}-\emph{specific}\xspace}
\newcommand\redout{\bgroup\markoverwith
{\textcolor{red}{\rule[0.5ex]{2pt}{0.8pt}}}\ULon}

\newboolean{showcomments}
\setboolean{showcomments}{true}

\ifthenelse{\boolean{showcomments}}
  {\newcommand{\nb}[2]{
    \fbox{\bfseries\sffamily\scriptsize#1}
    {\sf\small$\blacktriangleright$\textit{#2}$\blacktriangleleft$}
   }
  }
  {\newcommand{\nb}[2]{}
  }

\AtBeginDocument{%
  }

\newcommand{\apacheProjects}{1,161\xspace}
\newcommand{\apacheCommitsFilter}{1,114,142\xspace}
\newcommand{\apacheCommitsNoFilter}{1,272,556\xspace}

\newcommand{\springProjects}{68\xspace}
\newcommand{\springCommitsFilter}{74,906\xspace}
\newcommand{\springCommitsNoFilter}{84,591\xspace}

\newcommand{\trainedModels}{396\xspace}

\setcopyright{acmlicensed}
\copyrightyear{2025}
\acmYear{2025}
\acmDOI{XXXXXXX.XXXXXXX}

\begin{document}
	
	\title{Why Personalizing Deep Learning-Based Code Completion Tools Matters}
	
	\author{Alessandro Giagnorio}
	\affiliation{%
	  \institution{Universit\`{a} della Svizzera italiana}
	  \city{Lugano}
	  \country{Switzerland}}
	\email{alessandro.giagnorio@usi.ch}
	
	\author{Alberto Martin-Lopez}
	\affiliation{%
	  \institution{Universit\`{a} della Svizzera italiana}
	  \city{Lugano}
	  \country{Switzerland}}
	\email{alberto.martin@usi.ch}
	
	\author{Gabriele Bavota}
	\affiliation{%
	  \institution{Universit\`{a} della Svizzera italiana}
	  \city{Lugano}
	  \country{Switzerland}}
	\email{gabriele.bavota@usi.ch}
	
	\renewcommand{\shortauthors}{Giagnorio et al.}
	
	\begin{abstract}

		Deep learning (DL)-based code completion tools have revolutionized software development by providing unprecedented code generation capabilities. The DL models behind these tools are usually trained on large amounts of code from thousands of software repositories. This makes them good in learning \emph{natural} coding patterns observed across many training instances. However, little is known about the extent to which additional training effort (fine-tuning) aimed at specializing the models towards the code base of a given organization/developer further benefits their code completion capabilities. In this work, we fill this gap by presenting solid empirical evidence answering this question. More specifically, we consider 136 developers from two organizations (Apache and Spring), two model architectures (T5 and Code Llama), and three model sizes (60M, 750M, and 7B trainable parameters). For T5 models (60M, 750M), we pre-train and fine-tune them on over 2,000 open source projects, making sure that code from the two subject organizations is not part of their training sets. Then, we compare their completion capabilities against the same models further fine-tuned on organization- and developer-specific datasets. For the Code Llama model (7B), we compare the performance of the already pre-trained model publicly available online with the same model fine-tuned via parameter-efficient fine-tuning on organization- and developer-specific datasets. Our results show that there is a boost in prediction capabilities provided by both an organization-specific and a developer-specific  additional fine-tuning, with the former being particularly performant. Such a finding generalizes across (i) the two subject organizations (\ie Apache and Spring) and (ii) models of completely different magnitude (from 60M to 7B trainable parameters). Finally, we show that DL models fine-tuned on an organization-specific dataset achieve the same completion performance of pre-trained code models used out of the box and being $\sim$10$\times$ larger, with consequent savings in terms of deployment and inference cost (\eg smaller GPUs needed).
\end{abstract}

	\ccsdesc[500]{Software and its engineering}
	\ccsdesc[500]{Computing methodologies~Artificial Intelligence}
	
	\keywords{Software Engineering, Artificial Intelligence, Code Recommenders, Training Strategies}
	
	\received[revised]{DD MMMM YYYY}
	\received[accepted]{DD MMMM YYYY}
	
	
	\maketitle

\section{Introduction} \label{sec:intro}

The automatic generation of source code has been a long lasting dream in software engineering research for many years. Thanks to the advent of large deep learning (DL) models trained on code, we moved from predicting the next token the developer is likely to type \cite{Bruch:fse2009,Robb2010a} to the generation of complete code blocks and functions \cite{Ernst:sw2022}. Tools such as GitHub Copilot \cite{chen2021evaluating} are nowadays used by millions of developers and have been shown to boost their productivity~\cite{peng2023impact}.
These tools are trained on a large code corpora usually mined from open source projects. For example, Copilot's training set includes 159 GB of code mined from 54M public GitHub repositories \cite{chen2021evaluating}. Given the known repetitiveness of source code \cite{hindle:icse2012}, this training process allows the DL models to learn coding patterns seen across (possibly thousands of) code files, hence enabling the generation of meaningful recommendations when facing coding contexts similar to those in the training set. While the usefulness of these code recommenders is backed-up by empirical evidence \cite{peng2023impact}, there is still room for improvement when it comes to their performance\footnote{With performance we do not refer to properties such as execution time or memory usage, but to the accuracy of the generated recommendations.} \cite{Mastropaolo:icse2023}. 

One of the open questions when it comes to the adoption of DL-based code completion tools is whether their fine-tuning to the specific organization/developer using them may help in boosting performance. 
The idea is to perform a further training step after the ``generic fine-tuning'' (\ie the one in which the DL model is trained on code coming from thousands of repositories) with the aim of \emph{specializing} the DL model to a given code base (\eg the code developed within an organization or by a specific developer). Indeed, recent evidence from the Natural Language Processing (NLP) field demonstrated that more specific training data may help in boosting the performance of DL models, without risks of catastrophic forgetting their general knowledge. For example, Eschbach-Dymanus \etal \cite{eschbach-dymanus-etal-2024-exploring} showed that Large Language Models employed for natural language translation may benefit of additional fine-tuning aimed at specializing them for a specific domain. In our case, the ``specialized domain'' could be represented by the code base of a specific organization/company or by the code changes implemented over time by a single developer. Given the availability of open source DL models pre-trained on code (see \eg CodeBERT~\cite{feng2020codebert}, CodeT5~\cite{wang2021codet5}, Code Llama~\cite{roziere2023code}), showing the effectiveness of specializing them to a given code base may be relevant for companies who want to consider the possibility to fine-tune one of these models on their code base with the goal of deploying an in-house code recommender, possibly saving costs and avoiding potential issues related to the need to share proprietary code with a third-party DL model (\eg Copilot) to receive recommendations.

We present a large-scale empirical study investigating the extent to which personalizing DL-based code completion models helps in boosting their performance. We focus on two different levels of personalization, related to a whole software organization and a single developer. The former represents the scenario of a company specializing a single model on all software projects it runs. The latter answers the interesting research question of whether a deep level of personalization (down to the single developer) is really worthwhile. Indeed, a developer-specific personalization might be impractical, requiring the deployment and maintenance of several models. Still, if the gain in performance is major, then it could be considered in specific cases (\eg a small team).

From an abstract point of view, we start from a DL model $B$ that has been trained on a large and generic code corpus. $B$ represents our baseline, namely a ``generic'' DL-based code completion tool. We then collect code changes performed over time by developers who contributed to projects run by organization $org$. Given a developer $D$ who performed their contributions (\ie code changes, possibly spanning multiple projects run by $org$) over a time period $T_D$, we split $T_D$ into three parts, obtaining a $D$-\emph{specific} training, evaluation and test set. The test set features the most recent changes implemented by $D$. We then fine-tune $B$ on the $D$-\emph{specific} training set and compare its performance to our baseline ($B$) on the $D$-\emph{specific} test set. This shows the extent to which specializing a DL-based code completion tool to a specific developer $D$ improves the support provided to $D$ on future implementation tasks. Indeed, we are adopting a time-based splitting of data, ensuring that data from the past (the oldest $D$'s changes) is used to predict the future (the most recent $D$'s changes). Finally, we put together all previously built \emph{developer}-\emph{specific} training sets, thus creating an \emph{organization}-\emph{specific} training set. 
Such dataset has been used to specialize $B$ to the organization of interest ($org$), again comparing the performance of this specialized model to $B$. Also in this case the performance has been assessed on the \ds test sets, representing future changes that $org$'s developers will implement. 

Our study spans two organizations (Apache and Spring), two model architectures (T5~\cite{raffel2019exploring} and Code Llama~\cite{roziere2023code}) and three model sizes (60M, 750M and 7B trainable parameters). For T5 models, we pre-train and fine-tune them from scratch on a code base featuring over 2M instances (Java methods with some parts masked to simulate the code completion task). These models represent our baselines ($B$). Note that for these models we ensured that the training data used for the baselines did not include code from the organizations used as case studies (Apache and Spring). For the Code Llama model, this was not possible since the pre-trained model has been trained on a large code corpus which is not publicly available but it is \emph{very likely} to include code from both organizations. Still, it is interesting to observe if even in this case, a further personalized fine-tuning helps the model. In the case of Code Llama, the pre-trained model publicly available online represents our baseline.

Concerning the specialization of the models, we mine the change history of all Java projects hosted on GitHub by each organization, identifying the developers who contributed the most to these projects. To keep the experimentation affordable, we retrieve at most the top-100 developers (in terms of contributions) for each organization, provided that their contributions result in at least 1,000 training instances and 500 test instances, to make the training and evaluation meaningful. For Apache (Spring) we mined the change history of 1,161 (68) Java repositories, obtaining 100 (36) developers who contributed across all repositories. Throughout the document, we will continue to use the notation X (Y) to refer to the numbers related to Apache (Spring), respectively. Note that for Spring we only have 36 developers since the remaining ones did not meet the 1,000 training instances requirement. Following the previously-described process, we split their change history into three sets, ending up with 100 (36) \ds training, evaluation and test sets. We further fine-tune our $B$ baseline on these training sets, obtaining 100 (36) \emph{developer}-\emph{specialized} DL-based code completion models adopting the T5$_{small}$ architecture. We replicate the same process using the T5$_{large}$ and Code Llama models for the top-10 developers from each organization (20 in total). Each of the trained models has then been tested (and compared with the baseline) on the corresponding \emph{developer}-\emph{specific} test set. This analysis answers the question: \textbf{\emph{To what extent personalizing a DL-based code completion tool to the specific developer using it boosts its performance?}}

The \emph{organization}-\emph{specific} training sets have been obtained by merging the 100 (36) \emph{developer}-\emph{specific} datasets up to the latest date of the training set of each developer. This was done to ensure that data from the past is not used to predict the future (details in \secref{sec:design}). Thus, we created another 100 (36) \emph{organization}-\emph{specific} training sets, leading to 100 (36) new models based on the T5$_{small}$ architecture, plus 10 (for each organization) for the T5$_{large}$ and Code Llama models. This analysis answers the question: \textbf{\emph{To what extent personalizing a DL-based code completion tool to a software organization boosts its performance for individual developers of the organization?}}

On top of what described, we performed several analyses aimed at factoring out confounding variables, like ensuring that the observed improvement is not simply due to the additional data used for further fine-tuning the baselines. 

The above-summarized experiments required the training and testing of \trainedModels different models and showed that, while a very cheap fine-tuning performed on a developer-specific dataset boosts the performance of DL-based code completion tools, the observed improvement is usually capped by the limited number of developer-specific training data which can be collected for most developers. The organization-specific fine-tuning, instead, thanks to the additional training data available, works better than a developer-specific training, and should be the obvious choice in most of cases. Indeed, when considering both organizations (\ie Apache and Spring) together, the \os models achieve statistically significant improvements in correct predictions for: (i)~64\% of the 136 considered developers, with only one case of statistically significant drop in performance (T5$_{small}$); (ii)~70\% of the 20 considered developers, with no cases of significant drop in performance (T5$_{large}$); and (iii)~55\% of the 20 considered developers, again with no cases of significant performance drop (Code Llama). Also, we demonstrate that the increase in performance observed with both specializations is not simply due to a higher number of training instances as compared to the baselines, but to their specificity. Finally, through a cost-effectiveness analysis, we show that thanks to a personalized fine-tuning, DL models can achieve code completion performance on par with those of models being $\sim$10$\times$ larger (\eg an \os T5$_{small}$ achieves the same performance of a ``generic'' T5$_{large}$ model), possibly saving costs in the long-run.

The remainder of this paper is organized as follows: Section~\ref{sec:design} details the design of our study, including data collection, model training and evaluation. Section~\ref{sec:results} analyzes the results obtained across different levels of personalization and model sizes, while \secref{sec:findings} summarizes the main findings. Section~\ref{sec:threats} explains the validity threats and how these were mitigated. Finally, Section~\ref{sec:related} discusses related work, while Section~\ref{sec:conclusion} concludes the paper.

\section{Study Design} \label{sec:design}

We aim at answering the following research question: 

\begin{tcolorbox}[colback=gray!25, colframe=black!50, boxsep=5pt, arc=0pt, outer arc=0pt, boxrule=0.5pt, left=3pt, right=3pt, top=3pt, bottom=3pt]
    \begin{center}
    \textit{To what extent personalizing a DL-based code completion tool can boost its performance?}
    \end{center}
\end{tcolorbox}

We tackle this question by looking at the two levels of personalization previously mentioned: \ds and \os. The context of our study is represented by the two organizations considered, \ie Apache and Spring, and the code changes pushed by their top contributors to their Java projects hosted on GitHub. \tabref{tab:summary} summarizes the models, datasets, and metrics used in each part of our empirical investigation. We will refer to the goals described in the table (Goal X) throughout the section.

\begingroup
\definecolor{blue}{HTML}{008ED7}
\definecolor{mygray}{gray}{0.75}
\definecolor{lightBlue}{HTML}{e5f7ff}
\definecolor{darkGray}{gray}{0.2}
\definecolor{lightGray}{gray}{0.9}
\renewcommand{\familydefault}{\sfdefault}
\renewcommand{\arraystretch}{1.5}
\begin{center}
\fontfamily{cmss}\selectfont
\begin{table}[ht]
\caption{Summary of models, datasets, and metrics used in this study.\vspace{-0.3cm}}
\label{tab:summary}
\resizebox*{!}{0.95\textheight}{
\begin{tabularx}{\textwidth}[t]{XXX}

\arrayrulecolor{darkGray}\hline
\rowcolor{lightGray} \multicolumn{3}{l}{\textbf{\textcolor{darkGray}{Baselines' Training}}} \\
\hline
\small{\textbf{Models}} & \small{\textbf{Datasets}} & \small{\textbf{Metrics}} \\
\hline
\small
2 generic code-completion T5$_{small}$

\noindent\footnotesize
(1 Apache and 1 Spring)

\normalsize
\vspace{0.2cm}
\noindent
\small
2 generic code-completion T5$_{large}$

\noindent\footnotesize
(1 Apache and 1 Spring) & 
\normalsize
\small
2 pre-training datasets

\noindent\footnotesize
(1 Apache and 1 Spring)

\normalsize
\vspace{0.2cm}
\noindent
\small
2 code-completion datasets 

\noindent\footnotesize
(1 Apache and 1 Spring)
\normalsize & 
\small
Exact Match 

\noindent\footnotesize
(for best-checkpoint selection)\normalsize
 \\

\arrayrulecolor{mygray}\hline
\arrayrulecolor{blue}\hline
\rowcolor{lightBlue} \multicolumn{3}{l}{\textbf{\textcolor{blue}{Goal 1. Evaluating Developer- and Organization-Specific Personalization}}} \\
\hline
\small{\textbf{Models}} & \small{\textbf{Datasets}} & \small{\textbf{Metrics}} \\
\hline

\small
136 \ds T5$_{small}$ 

\noindent\footnotesize
(100 Apache and 36 Spring)
\normalsize

\vspace{0.2cm}
\noindent
\small
136 \os T5$_{small}$ 

\noindent\footnotesize
(100 Apache and 36 Spring) & 
\normalsize
\small
136 \ds train, evaluation, and test datasets 

\noindent\footnotesize
(100 Apache and 36 Spring)\normalsize

\vspace{0.2cm}
\noindent\small
136 \os train and evaluation datasets 

\noindent\footnotesize
(100 Apache and 36 Spring)\normalsize & 
\small
Exact Match and CrystalBLEU\normalsize\\
\hline
\rowcolor{lightBlue} \multicolumn{3}{l}{\textbf{\textcolor{blue}{Goal 2. Assessing the Impact of the Training Data Size}}} \\
\hline
\small{\textbf{Models}} & \small{\textbf{Datasets}} & \small{\textbf{Metrics}} \\
\hline

\small
20 \textit{Organization subset} T5$_{small}$ 

\noindent\footnotesize
(10 Apache and 10 Spring)

\normalsize
\vspace{0.2cm}
\noindent\small
20 \textit{Baseline+} T5$_{small}$ 

\noindent\footnotesize
(10 Apache and 10 Spring) & 
\normalsize

\small
20 \textit{Organization subset} train and evaluation datasets 

\noindent\footnotesize
(top 10 Apache contributors and 10 Spring contributors)\normalsize

\vspace{0.2cm}
\noindent\small
20 \textit{Baseline+} train and evaluation datasets

\noindent
\footnotesize
(top 10 Apache contributors and 10 Spring contributors)\normalsize & 
\small
Exact Match and CrystalBLEU\normalsize \\

\arrayrulecolor{mygray}\hline
\arrayrulecolor{blue}\hline
\rowcolor{lightBlue} \multicolumn{3}{l}{\textbf{\textcolor{blue}{Goal 3. Evaluating the Impact of Model Size, Architecture, and Pre-training}}} \\
\hline
\small{\textbf{Models}} & \small{\textbf{Datasets}} & \small{\textbf{Metrics}} \\
\hline

\small
20 \ds T5$_{large}$  

\noindent\footnotesize
(10 Apache and 10 Spring)
\normalsize

\noindent\small
20 \os T5$_{large}$ 

\noindent\footnotesize
(10 Apache and 10 Spring)

\normalsize
\vspace{0.2cm}
\noindent\small
20 \ds Code Llama

\noindent\footnotesize
(10 Apache and 10 Spring)

\normalsize
\noindent\small
20 \os Code Llama
\footnotesize(10 Apache and 10 Spring)\normalsize & 

\small
20 \ds train, evaluation, and test datasets from \textit{Goal 1}

\noindent
\footnotesize{(top 10 Apache contributors and 10 Spring contributors)}
\normalsize

\vspace{0.2cm}
\noindent\small
20 \os train and evaluation datasets from \textit{Goal 1}

\noindent
\footnotesize{(top 10 Apache contributors and 10 Spring contributors)}
& 
\small
Exact Match and CrystalBLEU \normalsize \\
\arrayrulecolor{mygray}\hline
\arrayrulecolor{blue}\hline
\rowcolor{lightBlue} \multicolumn{3}{l}{\textbf{\textcolor{blue}{Goal 4. Investigating the Cost-Performance Trade-Off}}} \\
\hline
\small{\textbf{Models}} & \small{\textbf{Datasets}} & \small{\textbf{Metrics}} \\
\hline

\small
2 generic code-completion T5$_{large}$ from \textit{Baselines’ Training}

\noindent\footnotesize
(1 Apache and 1 Spring) 
\normalsize
\vspace{0.2cm}

\noindent\small
20 \os T5$_{small}$ 

from \textit{Goal 1}

\noindent\footnotesize
(10 Apache and 10 Spring) & 

\small
20 \ds test datasets from \textit{Goal 1}

\noindent
\footnotesize{(top 10 Apache contributors and 10 Spring contributors)}
\normalsize & 
\small
Exact Match\normalsize \\

\end{tabularx}
}
\vspace{-0.7cm}
\end{table}
\end{center}
\endgroup

\subsection{Deep Learning Models}
As representative of DL models, we adopt the T5 \cite{raffel2019exploring} and Code Llama \cite{roziere2023code}. T5 is a transformer model already used in the literature to automate several code-related tasks \cite{Tufano:icse2022,isha:icscc,Berabi:mlr,Mastropaolo:tse2022}, including code completion \cite{ciniselli:tse2021}. Raffel \etal \cite{raffel2019exploring} proposed several variants of T5, differing in number of trainable parameters. We adopt two of them: \emph{small} and \emph{large}. The former features $\sim$60M parameters, while the latter $\sim$750M. All experiments described in \secref{sub:procedure} have been performed using the T5$_{small}$, while a subset of them features the T5$_{large}$ (as specified in \secref{sub:procedure}). Indeed, the trained \emph{large} variants allow us, in combination with Code Llama, to investigate whether the differences observed in terms of performance after personalizing the models are valid independently from the models' size. For both T5 variants, we use the T5v1.1 implementation available via Hugging Face \cite{T5imp}. During all trainings, the batch size is set to 32 for T5$_{small}$ and 4 for T5$_{large}$ (due to its higher cost in terms of GPU memory). For both models, we adopt their default hyper-parameter values, \ie learning rate of $5 \times 10^{-5}$, AdamW optimizer \cite{loshchilov:iclr19} and linear decay scheduler.

Code Llama \cite{roziere2023code} has also been proven to be effective on a broad range of Software Engineering tasks \cite{weyssow2023exploring, cassano2024knowledge, huang2024template, sun2024source}. Code Llama is a family of transformer models based on the general-purpose Llama 2 \cite{touvron2023llama} and further trained on 500B code-specific data tokens. It comes in different sizes (7B, 13B, 34B, and 70B) and versions (base model, instruction-tuned, and Python specialist). We select the base variant with 7B parameters, which is $\sim$10 times larger than the T5$_{large}$ model. Adding this model to our study allows us to: (i) compare the performance of the T5 models with state-of-the-art code models like Code Llama; (ii) investigate the impact of personalization on models with a very large number (\ie billions) of parameters; (iii) understand the generalizability of our findings to a different model architecture; and (iv)~assess the impact of personalization on a pre-trained model, whose training dataset may already feature code from the organization and developers of interest. To reduce training costs for this larger model, we train Code Llama with the LoRA technique~\cite{hu2022lora}. LoRA is a Parameter-Efficient Fine-Tuning (PEFT) method that aims to approach the performance of full-parameter fine-tuning by only training a small number of parameters. This technique freezes the pre-trained weights and replaces the gradient update matrix with two trainable low-rank matrices. This significantly reduces the number of trainable parameters (\eg from 7B to 40M for Code Llama) while also lowering the computational cost. Following a previous study on code generation \cite{weyssow2023exploring}, we set the LoRA hyperparameter values to $\alpha = 32$ and $r=16$, while using the same training configuration seen for T5$_{large}$.

\subsection{Datasets Used for Training and Testing}
\label{sub:datasets}

We describe the datasets used to train the DL models for the code completion task. In such a context, a training instance is a Java method having some contiguous tokens masked, with the model in charge of predicting them. Indeed, as done in previous work on DL-based code completion \cite{ciniselli:tse2021}, we work at method-level granularity. We start by describing the \ds (\secref{sub:dev}) and the \os (\secref{sub:org}) datasets. Then, we present our generic pre-training and fine-tuning datasets used to train the T5 models (\secref{sub:generic-datasets}), which are not already trained like Code Llama. 

\subsubsection{Developer-Specific Datasets}
\label{sub:dev}
\begin{figure}[t]
    \centering
    \includegraphics[width=1\columnwidth]{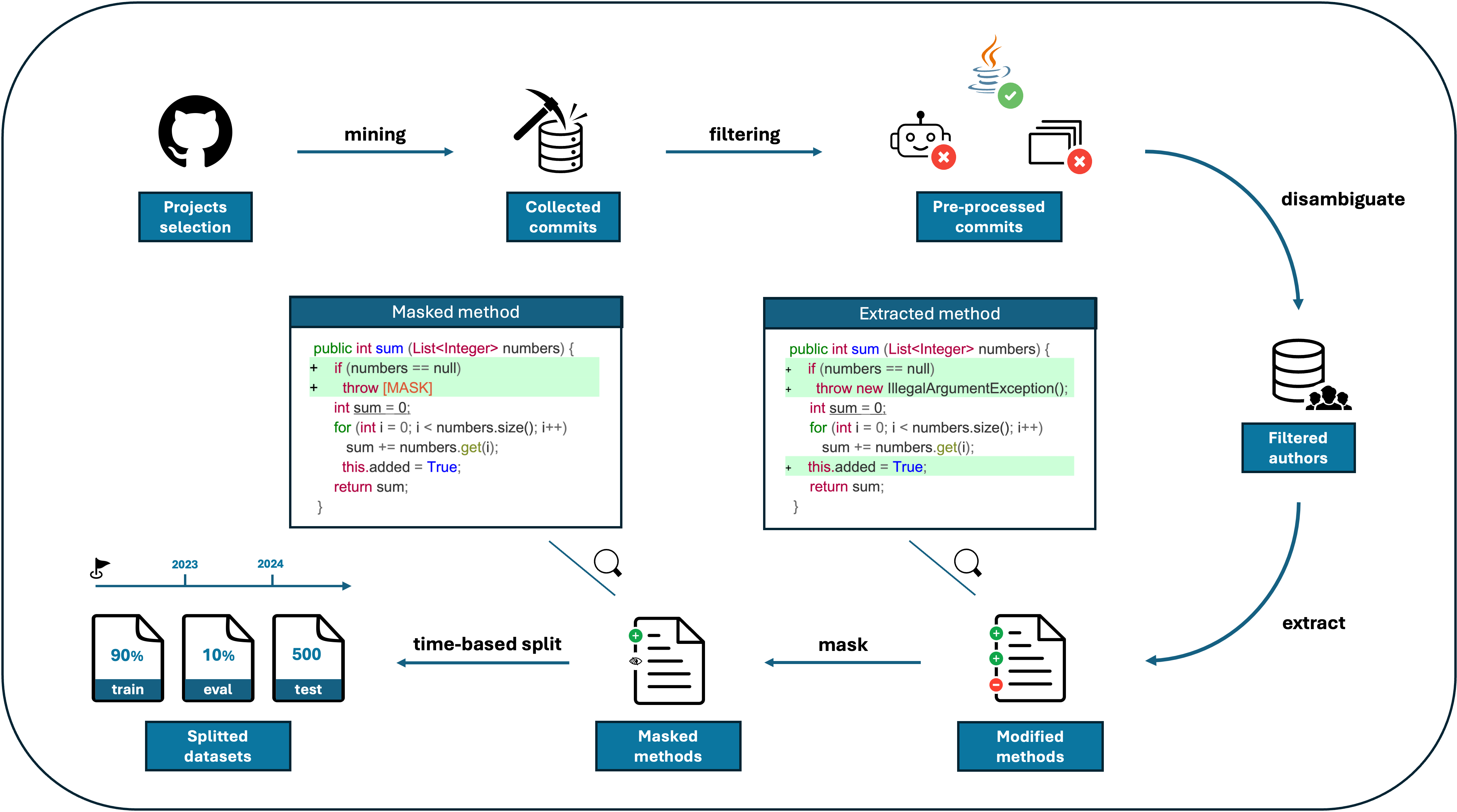}
    \caption{Mining process to create the \ds datasets.}
    \label{fig:mining-process}
\end{figure}

For both the \emph{developer}- and \emph{organization}-\emph{specific} datasets, our goal is to create training/testing instances that are representative of real code changes performed by developers belonging to the organization of interest (\ie Apache or Spring). \figref{fig:mining-process} illustrates the process we followed to create the \ds datasets. We explain it in the following.

\textbf{Commits mining.}
We start by mining all commits from the main branch of the \apacheProjects Apache (\springProjects Spring) Java projects considered in our study. We exclude commits performed by bots, not modifying Java files, and impacting too many files, likely being the result of automated operations (\eg initial commit, rename package refactoring, \etc). Concerning the identification of bots, we apply a simple heuristic filtering out all commits performed by  authors having a name containing ``[bot]'' and/or ``GitHub''. As for the commits impacting a large number of files, once mined all commits, we exclude those being outliers in terms of number of modified files, \ie impacting more than $Q_3 + 1.5 \times IQR$ files, where $Q_3$ is the third quartile and $IQR$ is the interquartile range of the distribution of impacted files across all commits. This process narrowed down the initial set of \apacheCommitsNoFilter (\springCommitsNoFilter) commits to \apacheCommitsFilter (\springCommitsFilter) relevant commits. 

\textbf{Extracting Java methods featuring new code.}
The next step consists in parsing the Java files impacted in the mined commits with the goal of identifying Java methods in which at least one new line has been added. We only focus on added lines and ignore the modified ones since our idea is to exploit these methods to generate training instances in which the code written by a specific developer (\ie the added lines) is masked, and the model is in charge of predicting it. For modified lines, in theory, a developer may change a single token (or even a space) and it would be wrong to assume that the modified line represents code written by the developer. We expect this process to specialize the model towards the code changes representative of the work done by a developer and, thus, of the software organization they contribute to. 

Since parsing the Java files associated with each commit is costly, we decided to perform this process only for the top-1k developers (in each organization) in terms of added lines of code. Those are the ones likely to provide enough ``specialized training data'' which can then be used to experiment with the \ds fine-tuning. Selecting these top-1k developers is not trivial. Indeed: (i) the developers who authored the mined commits could have contributed to more than one of the \apacheProjects Apache (\springProjects Spring) projects we mine, possibly using different names/emails; (ii) the change history associated with several of the subject repositories is extremely long (\eg $>$20 years for Apache Commons BeanUtils~\cite{apache-beanutils}), again increasing the likelihood that a developer changed name/email used when committing to the versioning system over time. For these reasons, before selecting the top-1k developers, we use the \texttt{gambit} tool~\cite{gote2021gambit} to disambiguate the authors of all commits, associating the same \emph{author ID} to commits performed by the same developer using different names. Once identified the top-1k contributors, we manually inspect the disambiguations, excluding wrong matches. In particular, given an author for which multiple GitHub accounts were matched, the first author inspected all of them discarding cases in which (i) the account cannot be traced back to the same person with high confidence; or (ii) one or more of the matched accounts does not exist anymore. This process left us with 686 (818) valid and disambiguated developers. 

For each commit they performed, we clone the corresponding repository at the commit's hash and use \texttt{javalang}~\cite{javalang} to parse the impacted Java files and extract all methods with at least one line added (according to the commit's diff). We subsequently discard methods that: (i)~cannot be parsed (\eg they contain syntax errors); (ii)~contain the word ``test'' in their name (after splitting it \emph{camelCase}-wise), to create a more coherent dataset of production code; (iii)~contain an empty body or only comments; (iv)~contain less than 15 (too simple) or more than 500 tokens (too long to handle with the subject DL models); and (v)~contain non-latin characters (\eg emojis). Through these filters, we obtained 1,148,324 (197,622) Java methods with at least one new line implemented in a given commit.

\textbf{Creating training, evaluation and test sets for the top developers.}
As a final step, we create training/testing instances from each method extracted in the previous steps. We indicate with $L$ all lines added in a method in a specific commit. As a running example, let us assume that $L=\{l_4$, $l_5$, $l_6$, $l_7$, $l_8$, $l_{14}$\}, with the subscript number indicating the line number of each added line.
If a new line is added in ``isolation'' (\ie it does not have other added lines right before/after it), we mask its last $n$ tokens with $n$ randomly ranging from 3 to $min(50,$ $N - 1)$, where $N$ is the total number of tokens in the line. This is what happens in our running example to line $l_{14}$. Note that we mask at least 3 and at most 50 tokens to avoid trivial completions (\eg predicting only the last semicolon of a statement) while keeping the completion task approachable (max. 50 tokens to predict). If, instead, a block of contiguous lines is added ($l_4$ to $l_8$ in our example), we split it into blocks of at most three lines, with empty lines (\ie those added for formatting purpose) or lines featuring a single token (\eg `\texttt{\}}') not counting towards this limit. In our example, $l_6$ is an empty line, therefore $l_4$ to $l_7$ becomes one block, and $l_8$ a second block. We then apply the same masking procedure described for the isolated lines. This means that we compute $N$ as the total number of tokens in the block, and mask its last $n$ tokens with $n$ randomly selected as previously described. Again, we limit the maximum number of lines in a block to three to keep the complexity of the completion task reasonable. Two important points are worth being noticed. First, both completion tasks (\ie masking of a single line or of a block of lines) simulate a developer that starts writing the needed code and receives support to complete it, with the recommendation possibly featuring multiple statements in the case of block completion. Second, a single method featuring multiple added lines in a commit can contribute multiple training/testing instances, each featuring different added line(s) masked. 

The output of  the aforementioned process is a dataset of code completion instances assigned to each of the 686 (818) developers, with a number of instances ranging from 42 (0) to 52,638 (20,358). For each developer, we order their dataset chronologically and keep the most recent 500 instances (code additions) as test set, while splitting the rest in the 90\% least recent instances for training and the 10\% most recent instances for validation. Following this procedure we make sure that only data from the past is used to predict the future, resembling a real-world scenario. Duplicates across the training and the evaluation/test sets are removed from the training set. After this process, we keep up to 100 developers from each organization, provided that they feature at least 1,000 training instances. This leads to 100 developers for Apache and 36 for Spring.

\subsubsection{Organization-Specific Datasets}
\label{sub:org}
We create the \os datasets by exploiting the code completion instances previously created for the top developers. The idea is that instances crafted starting from the code changes of the top developers are representative of the implementation tasks usually carried out in the organization. Since also the \os models will be tested on the \ds test sets---100 (36) different test sets---we must create 100 (36) different \os training datasets, to make sure that all code changes used to build the \os training instances are older than those used to build the (\ds) testing instances.

\begin{figure}[ht]
    \centering
    \includegraphics[width=0.5\columnwidth]{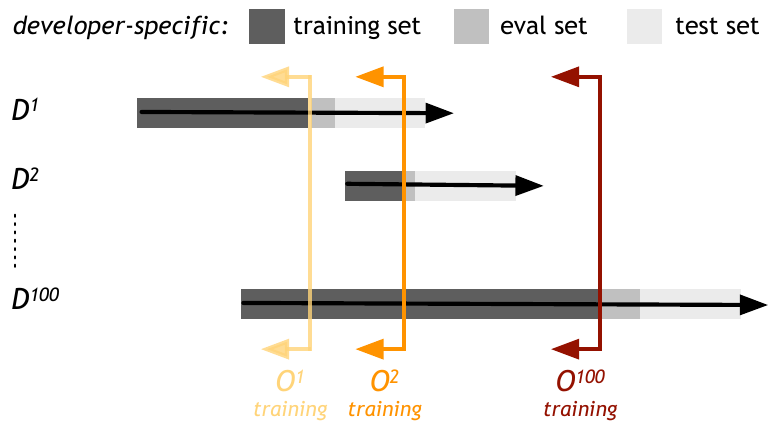}
    \caption{Developer- and organization-specific datasets.}
    \label{fig:datasets}
\end{figure}

\figref{fig:datasets} depicts the process to create the 100 \os datasets for the Apache organization, one per each \ds test set. The same process has been used for Spring, starting from the 36 \ds datasets. The 100 \ds datasets are represented in \figref{fig:datasets} using the $D^1 \dots~D^{100}$ notation. The arrow associated to each developer represents the history of their changes, possibly spanning across multiple repositories (older changes to the left). As explained in \secref{sub:dev}, for the \ds dataset the 500 code completion instances derived from the most recent commits are taken as test set, while the rest is split into 90\% training and 10\% validation. The history of changes is not aligned among the 100 developers. This means, for example, that we have developers who started contributing in 2010 while others who started contributing in 2020. This is the reason why the arrows in \figref{fig:datasets} are not aligned. When creating the \os training set for the model that will be tested on $D^1$'s test set (see the yellow arrow labeled with ``$O^1$ training''), we take the code completion instances derived from the changes performed by all 100 developers up to the date of the most recent change in $D^1$'s training set (see \figref{fig:datasets}). In this way, we ensure that the $O^1$ training set only features instances related to changes older than both $D^1$'s validation and test set.

\subsection{Generic Pre-training and Fine-tuning Datasets}
\label{sub:generic-datasets}
The datasets previously described are used to specialize (\ie fine-tune) the DL models towards the organization/developer of interest. While Code Llama is a pre-trained model that can be trained with such datasets, the T5 models call for a pre-training and generic fine-tuning phase. This section describes the datasets used for these purposes.

We start from the CodeParrot GitHub Code dataset~\cite{codeparrot}, featuring 19.5M Java files, 108GB of data, and including metadata about the repositories from which the code was mined. We create two subsets of it, one excluding all instances mined from Apache repositories and one all those from Spring repositories, to simulate a scenario where the generic DL-based code completion tool is not personalized towards the organization of interest (\eg the code of the organization is not publicly available). Also, we exclude all repositories being forks of others and not having at least 100 commits, 10 contributors and 10 stars. We do this in an attempt to remove toy/personal projects. The dataset from which we excluded Apache repositories featured, after this cleaning, 2,098 projects, while the one from which we excluded Spring repositories was left with 2,057 projects. In both cases, those have been split into 40\% for pre-training and 60\% for fine-tuning.
As done for the \ds datasets, we generate training instances at method-level granularity and apply the same filters previously described (\eg remove test methods, too short/long methods, \etc). For each repository, we randomly collect at most 1,500 valid methods, to avoid biasing the dataset towards repositories with a large number of methods.

\subsubsection{Pre-training Dataset}
\label{sub:pretraining}
For pre-training, we collect a total of 1,142,330 (1,091,327) methods. We use the masked language modelling training objective~\cite{devlin:naacl-hlt2019}, randomly masking 15\% of tokens of the input method and asking the model to predict them. We split the dataset into 90\% for training and 10\% for validation.

\subsubsection{Fine-tuning Dataset}
\label{sub:finetuning}
After removing duplicates with all \ds test sets, we end up with 1,080,909 (1,355,885) methods for fine-tuning. To have a fair comparison between the specialized DL models and the generic ones (\ie the ones trained on a large dataset not featuring code coming from the organization of interest), we aim to create a generic training dataset that is as similar as possible to the \ds ones and, as a consequence, to the \os one as well. To this end, we compute in the \ds datasets (all merged together) the distribution of the number of tokens masked per instance (mean=11, median=8, min=3, max=50 for Apache and mean=13, median=10, min=3, max=50 for Spring). 
We tried to mirror this distribution when generating the fine-tuning instances for the generic dataset by randomly masking the end tokens of lines/blocks in its methods. We generate at most three code completion instances per method (\ie three versions of the same method with different parts masked), obtaining 1,434,598 (1,355,862) code completion instances, which we split into 90\% for training and 10\% for validation. These instances have a mean of 11 (13) masked tokens, with a median of 8 (10), minimum of 3 (3) and maximum of 50 (50) tokens, resembling the characteristics of the \ds datasets.

\subsection{Experimental Procedure and Data Analysis}
\label{sub:procedure}
Code Llama is a pre-trained model that supports code completion out of the box, while T5 does not. Thus, we start explaining the process used to train the two T5 variants considered in our study.
We pre-train the T5 models on the dataset described in \secref{sub:pretraining} using early stopping, with checkpoints saved every epoch, a delta of 0.005, and a patience of 5 (Baselines’ Training in \tabref{tab:summary}). This means that the models are evaluated on the pre-training validation set every epoch in terms of percentage of correctly predicted masked tokens, and the training stops if a gain in performance lower than delta is observed at each 5-epoch interval. 
Once pre-trained, we fine-tune the T5 variants on the generic fine-tuning dataset described in \secref{sub:finetuning} (Baselines’ Training). We use the same early stopping procedure previously described for the pre-training with, however, a delta of 0.01, since we observed a faster increase in the models' prediction accuracy during fine-tuning (probably due to the fact that the models were already pre-trained). In this case, the performance of the models at each epoch has been assessed in terms of Exact Match (EM) predictions on the fine-tuning validation set (\ie the predicted code is identical to the masked one).

The pre-trained and fine-tuned T5 models (both small and large) and the already trained Code Llama model publicly available represent our baselines (\ie generic models trained on a large amount of code). We indicate these models with B$_{s}$ (T5$_{small}$), B$_{l}$ (T5$_{large}$) and B$_{c}$ (Code Llama).

We use the same early stopping procedure described for the fine-tuning to further train B$_{s}$, B$_{l}$, and B$_{c}$, and obtain their \ds and \os versions (Goal 1). For what concerns B$_{s}$, we further fine-tune 272 versions of it: 100 (36) \ds and 100 (36) \os versions. Indeed, as explained in \secref{sub:org}, even for the \os models we had to create 100 (36) different training sets to avoid using data from the future to predict the past. Note that these are 272 different models, all representing further fine-tunings of B$_{s}$. We use the notation D$^{i}_{s}$ to indicate the B$_{s}$ model fine-tuned on the \ds dataset of the developer $D^{i}$. Similarly, we denote with O$^{i}_{s}$ the B$_{s}$ model fine-tuned on the \os dataset temporally aligned to the training set of $D^{i}$ (\ie not including changes performed after the last change in $D^{i}$'s training set). 

We compare the performance of both D$^{i}_{s}$ and O$^{i}_{s}$ against the baseline (B$_{s}$) on $D^{i}$'s test set (Goal 1), since we want to verify whether the $D^{i}$-\emph{specific} and the \os models can better predict future $D^{i}$'s changes as compared to a generic code completion model (B$_{s}$). As evaluation metric, we compute the percentage of EM  predictions in the test set. While this metric has been used in several code completion works \cite{alon:icml2020,asaduzzaman:icsme2014,Ciniselli:msr2021,ciniselli:tse2021,hellendoorn:fse2017}, it represents a lower bound for the performance of a given approach. Indeed, it considers a prediction as correct only if the generated code is identical to the one to predict. This means that different but semantically equivalent code generated by the model will be considered wrong. For this reason, we complement our analysis by computing the CrystalBLEU score \cite{eghbali:icse2022} between the generated predictions and the expected code. CrystalBLEU is a variant of the BLEU score~\cite{papineni:acl2002} tailored for code and has been shown to correctly discriminate similar from dissimilar code 1.9--4.5 times more effectively when compared to BLEU \cite{eghbali:icse2022}. We statistically compare the results achieved by the different models. For the EM predictions, we use the McNemar's test \cite{mcnemar}, which is  suitable to pairwise compare dichotomous results of two different treatments. We complement the McNemar's test with the Odds Ratio (OR) effect size. As for the CrystalBLEU, we use the Wilcoxon signed-rank test \cite{wilcoxon} and the paired Cliff's delta \cite{Cliff:2005} effect size. 

We perform the same training procedure and data analysis using the B$_{l}$ (T5$_{large}$) and B$_{c}$ (Code Llama) baselines, thus obtaining specialized models D$^{i}_{l}$ and O$^{i}_{l}$, compared against B$_{l}$, and D$^{i}_{c}$ and O$^{i}_{c}$, compared against B$_{c}$ (Goal 3). As said, such an analysis has been performed only for the top-10 developers of each organization in terms of contributed code changes, thus obtaining 40 additional T5$_{large}$ and 40 additional Code Llama models---10 Apache (10 Spring) \ds and 10 Apache (10 Spring) \os.

\subsubsection{Controlling for the Training Data Size Confounding Factor}
\label{sec:controlling-size}
We study the impact of the amount of training provided to the models on their performance (Goal 2): since we found that the \os models work better, there is a question related to the extent to which this is due to the additional training data it has seen as compared to the baselines. Indeed, since the \os models are obtained via further fine-tuning the baseline, they benefit from more training data. To control for such a confounding factor, we further fine-tune B$_{s}$ in two different ways: (i)~with an \os dataset capped to the same size of the \ds dataset, leading to \emph{Organization subset} models, and (ii)~with a generic dataset capped to the same size of the \os dataset, leading to \emph{Baseline+} models. By comparing the performance of \ds with \emph{Organization subset} models, since both have seen the same amount of training instances, we can understand the impact of the \ds data on the models' performance. Similarly, by comparing the performance of \os with \emph{Baseline+} models, we can understand the impact of the \os data on the models' performance. To create the generic dataset (used to fine-tune the \emph{Baseline+} models), we mined 2,354 (2,781) additional Java open source repositories hosted on GitHub which are not from Apache or Spring---thus avoiding overlap with the \os datasets. These repositories have been collected using the platform by Dabi\'c \etal \cite{dabic:msr2021}, querying it for all Java repositories having at least 10 contributors, 10 stars, and 100 commits. We processed them using the same procedure previously described for the ``generic fine-tuning'' dataset (\secref{sub:finetuning}).

In practice, we created 20 ``organization subset'' training sets and 20 ``generic'' training sets, one for each of the top-10 developers of both organizations. This was done to ensure that the training sets associated to a developer $D$ would only contain instances whose date was before the first date of $D$'s test set, \ie to avoid using data from the future to predict the past. Due to the further training cost, these analyses have been performed only with the T5$_{small}$ model.

\subsubsection{Investigating Why More Specific Training Data Help}
\label{sec:specific-data}
Since our findings show that more specific training data help the model, we also perform an additional analysis aimed at investigating the \emph{information items} shared between the three training sets (\ie the ones used by the baseline B$_s$, by the \ds, and by the \os) and the 136 developer-related test sets of the two organizations. With information items we refer to:

\begin{itemize}
\item \emph{Domain coverage}: The extent to which the domain of the data present in the test sets is covered in the training sets. \emph{Method signatures} are a good proxy to this end, since they include information about supported operations (\eg via method names) and input/output data (\eg via return types and method arguments). We thus compute the percentage of instances in the test sets whose method signature appears in the training sets.
\item \emph{Vocabulary coverage}: The extent to which the vocabulary used in the test sets is covered in the training sets. We focus on literals (\eg strings and numbers) and identifiers (\eg method and variable/constant names), reporting the percentage of these elements in the test sets which are also present in the training sets.
\item \emph{Relevance of the Training Data}: The extent to which the \emph{vocabulary learned during training} is actually used at inference time (\ie present in the test sets). We compute the percentage of identifiers and literals in the training sets which can be found in the test sets.
\end{itemize}

\subsubsection{Cost-effectiveness Analysis}
\label{sec:cost-effectiveness}
We also run an additional analysis aimed at understanding the cost-effectiveness of the ``personalized'' fine-tuning (Goal 4). Indeed, the personalized fine-tuning implies a training cost that the company would not have by just downloading a larger code-completion model already trained for such a task, and maybe exhibiting even better performance than the smaller, personalized model. We perform this cost-effectiveness analysis between T5$_{small}$ with ``personalized fine-tuning'' and T5$_{large}$ (being 12.5 times bigger) with a generic fine-tuning, as representative of an already generic trained model which can be downloaded and used with \emph{zero} training cost. We excluded Code Llama from this part of the study since in our setting it makes no sense to consider Code Llama as representative of a general-purpose model that a company could download and use out of the box, since it likely saw the code of the two organizations (Apache and Spring) subject of our study during training. Thus, any sort of cost-effectiveness analysis comparing a non fine-tuned Code Llama versus a fine-tuned and personalized T5 would not allow to observe the actual advantage (if any) provided by personalization. Instead, by considering the T5$_{large}$ pre-trained and fine-tuned on a generic dataset as an example of trained model that a company can just download and use out of the box, we can guarantee that (i) this is a model that has not seen the company's code, since we excluded that code from the pre-training and fine-tuning datasets; and (ii) we are still considering a model that is 12.5 times bigger than T5$_{small}$, thus allowing to observe if a much smaller model with personalized fine-tuning is able to reach similar performance of a much larger (non personalized) model. Since the T5$_{large}$ has only been assessed on the top-10 developers of each of the two organizations, this cost-effectiveness analysis has been performed on these 20 developers. Also, among the two personalizations of T5$_{small}$ (\ie developer-specific and organization-specific), we considered the organization-specific which is more expensive (larger training sets) but achieves better performance (as we will show, aligned to those of T5$_{large}$).

To present reliable data we computed the cost of renting GPUs in Google Cloud, for both the fine-tuning of T5$_{small}$ and the inference performed with both models. We considered the training cost of both the cheapest (146.3k training instances) and the most expensive (888k training instances) organization-specific T5$_{small}$. For the training of T5$_{small}$ and the inference of both T5$_{small}$ and T5$_{large}$, we rented 1 Nvidia T4 GPU with 16GB of memory. To compute the inference cost, we generate the same 1,000 predictions with each model, and then compute the average cost of one prediction. Clearly, while the GPU used for inference is the same for all models, T5$_{large}$ requires a longer inference time (higher cost). We discuss the cost that an organization would have with both models given a different number of inferences performed by its developers, presenting break-even points in the best- (\ie lowest number of training instances for the organization-specific fine-tuning, 146.3k) and worst-case (highest number, 888k) scenario. In other words, we discuss after how many inferences the company would amortize the fine-tuning cost of the personalized smaller model and start saving money.
\section{Results Discussion} \label{sec:results}
\begin{sidewaystable*}
    \fontsize{12}{13}\selectfont
    \centering
    \vspace*{15cm}
    \caption{Exact Match (EM) predictions generated by the baseline and by the personalized models for Apache.}
    \label{tab:results-t5small-em}
    \vspace*{-0.1cm}
    \rowcolors{2}{gray!20}{white}
    \begin{minipage}{0.48\textwidth}
        \centering
        \scalebox{0.59}{
\begin{tabular}{|c|c|cccr|cccr|}
            \hiderowcolors
            \toprule 
            \multicolumn{1}{c}{\raisebox{-\heavyrulewidth}{\textbf{Dev. ID}}} & \multicolumn{1}{c}{\raisebox{-\heavyrulewidth}{\textbf{Baseline B$_{s}$}}} & \multicolumn{4}{c}{\textbf{Developer}} & \multicolumn{4}{c}{\textbf{Organization}}\\
            \cmidrule{1-10}
            & \textit{EM~\small\%} & \textit{N°} & \textit{EM~\small\%} & \textit{$\Delta$} & \textit{OR} & \textit{N°} & \textit{EM~\small\%} & \textit{$\Delta$} & \textit{OR} \\
            \showrowcolors
            \midrule
1 & 51.2 & 46.6k & 60.8 & \color{green}\raisebox{0ex}{$\blacktriangle$}~\textcolor{black}{\texttt{+}\hspace{2ex}9.60\small{\%}} & 6.33 & 888.0k & \textbf{61.8} & \color{green}\raisebox{0ex}{$\blacktriangle$}~\textcolor{black}{\texttt{+}\hspace{0.875ex}10.60\small{\%}} & 8.57 \\
2 & 31.8 & 23.2k & 33.2 & \color{gray!20}\raisebox{0ex}{$\blacktriangle$}~\textcolor{black}{\texttt{+}\hspace{2ex}1.40\small{\%}} & 1.28 & 556.0k & \textbf{37.0} & \color{green}\raisebox{0ex}{$\blacktriangle$}~\textcolor{black}{\texttt{+}\hspace{2ex}5.20\small{\%}} & 3.36 \\
3 & 26.0 & 20.3k & 25.0 & \color{white}\raisebox{0ex}{$\blacktriangledown$}~\textcolor{black}{\texttt{-}\hspace{2ex}1.00\small{\%}} & 0.62 & 830.8k & \textbf{28.2} & \color{white}\raisebox{0ex}{$\blacktriangle$}~\textcolor{black}{\texttt{+}\hspace{2ex}2.20\small{\%}} & 1.85 \\
4 & 35.2 & 19.2k & 39.0 & \color{green}\raisebox{0ex}{$\blacktriangle$}~\textcolor{black}{\texttt{+}\hspace{2ex}3.80\small{\%}} & 1.83 & 747.3k & \textbf{44.2} & \color{green}\raisebox{0ex}{$\blacktriangle$}~\textcolor{black}{\texttt{+}\hspace{2ex}9.00\small{\%}} & 3.65 \\
5 & 21.4 & 18.6k & 18.6 & \color{red}\raisebox{0ex}{$\blacktriangledown$}~\textcolor{black}{\texttt{-}\hspace{2ex}2.80\small{\%}} & 0.42 & 540.5k & \textbf{24.0} & \color{green}\raisebox{0ex}{$\blacktriangle$}~\textcolor{black}{\texttt{+}\hspace{2ex}2.60\small{\%}} & 2.30 \\
\bottomrule
        \end{tabular}
        }
    \end{minipage}
    \hfill
    \begin{minipage}{0.48\textwidth}
        \centering
        \scalebox{0.59}{
        \begin{tabular}{|c|c|cccr|cccr|}
            \hiderowcolors
            \toprule 
            \multicolumn{1}{c}{\raisebox{-\heavyrulewidth}{\textbf{Dev. ID}}} & \multicolumn{1}{c}{\raisebox{-\heavyrulewidth}{\textbf{Baseline B$_{s}$}}} & \multicolumn{4}{c}{\textbf{Developer}} & \multicolumn{4}{c}{\textbf{Organization}}\\
            \cmidrule{1-10}
            & \textit{EM~\small\%} & \textit{N°} & \textit{EM~\small\%} & \textit{$\Delta$} & \textit{OR} & \textit{N°} & \textit{EM~\small\%} & \textit{$\Delta$} & \textit{OR} \\
            \showrowcolors
            \midrule
51 & 26.8 & 5.6k & 26.0 & \color{white}\raisebox{0ex}{$\blacktriangledown$}~\textcolor{black}{\texttt{-}\hspace{2ex}0.80\small{\%}} & 0.73 & 877.8k & \textbf{29.8} & \color{green}\raisebox{0ex}{$\blacktriangle$}~\textcolor{black}{\texttt{+}\hspace{2ex}3.00\small{\%}} & 2.25 \\
52 & 28.0 & 5.5k & \textbf{29.0} & \color{gray!20}\raisebox{0ex}{$\blacktriangle$}~\textcolor{black}{\texttt{+}\hspace{2ex}1.00\small{\%}} & 1.31 & 408.3k & \textbf{29.0} & \color{gray!20}\raisebox{0ex}{$\blacktriangle$}~\textcolor{black}{\texttt{+}\hspace{2ex}1.00\small{\%}} & 1.23 \\
53 & 32.8 & 5.5k & 35.2 & \color{white}\raisebox{0ex}{$\blacktriangle$}~\textcolor{black}{\texttt{+}\hspace{2ex}2.40\small{\%}} & 1.41 & 846.2k & \textbf{50.8} & \color{green}\raisebox{0ex}{$\blacktriangle$}~\textcolor{black}{\texttt{+}\hspace{0.875ex}18.00\small{\%}} & 12.25 \\
54 & 29.2 & 5.5k & 30.6 & \color{gray!20}\raisebox{0ex}{$\blacktriangle$}~\textcolor{black}{\texttt{+}\hspace{2ex}1.40\small{\%}} & 1.64 & 211.0k & \textbf{33.2} & \color{green}\raisebox{0ex}{$\blacktriangle$}~\textcolor{black}{\texttt{+}\hspace{2ex}4.00\small{\%}} & 2.82 \\
55 & 29.4 & 5.3k & 29.6 & \color{white}\raisebox{0ex}{$\blacktriangle$}~\textcolor{black}{\texttt{+}\hspace{2ex}0.20\small{\%}} & 1.08 & 368.6k & \textbf{32.8} & \color{green}\raisebox{0ex}{$\blacktriangle$}~\textcolor{black}{\texttt{+}\hspace{2ex}3.40\small{\%}} & 2.31 \\
\bottomrule
        \end{tabular}
        }
    \end{minipage}
\end{sidewaystable*}
\normalsize
\begin{sidewaystable*}
    \fontsize{12}{13}\selectfont
    \centering
    \vspace*{15cm}
    \caption{CrystalBLEU (CB) average score between the baseline and the personalized models for Apache.}
    \label{tab:results-t5small-cb}
    \vspace*{-0.1cm}
    \rowcolors{2}{gray!20}{white}
    \begin{minipage}{0.48\textwidth}
        \centering
        \scalebox{0.62}{
        \begin{tabular}{|c|cccr|cccr|}
            \hiderowcolors
            \toprule 
            \multicolumn{1}{c}{\raisebox{-\heavyrulewidth}{\textbf{~Dev. ID~}}} & \multicolumn{4}{c}{\textbf{Developer}} & \multicolumn{4}{c}{\textbf{Organization}}\\
            \cmidrule{1-9}
            & \textit{N°} & \textit{CB~\small\%} & \textit{$\Delta$} & \textit{$\vert ES \vert$} & \textit{N°} & \textit{CB~\small\%} & \textit{$\Delta$} & \textit{$\vert ES \vert$} \\
            \showrowcolors
            \midrule
1 & 46.6k & 38.24 & \color{green}\raisebox{0ex}{$\blacktriangle$}~\textcolor{black}{\texttt{+}\hspace{0.875ex}18.77\small{\%}} & 0.23 & 888.0k & {41.28} & \color{green}\raisebox{0ex}{$\blacktriangle$}~\textcolor{black}{\texttt{+}\hspace{0.875ex}22.45\small{\%}} & 0.29 \\
2 & 23.2k & 28.75 & \color{gray!20}\raisebox{0ex}{$\blacktriangle$}~\textcolor{black}{\texttt{+}\hspace{2ex}3.63\small{\%}} & 0.06 & 556.0k & {30.82} & \color{green}\raisebox{0ex}{$\blacktriangle$}~\textcolor{black}{\texttt{+}\hspace{2ex}8.68\small{\%}} & 0.13 \\
3 & 20.3k & 19.82 & \color{white}\raisebox{0ex}{$\blacktriangledown$}~\textcolor{black}{\texttt{-}\hspace{2ex}1.51\small{\%}} & 0.01 & 830.8k & {26.70} & \color{green}\raisebox{0ex}{$\blacktriangle$}~\textcolor{black}{\texttt{+}\hspace{2ex}5.37\small{\%}} & 0.07 \\
4 & 19.2k & 28.07 & \color{green}\raisebox{0ex}{$\blacktriangle$}~\textcolor{black}{\texttt{+}\hspace{2ex}5.59\small{\%}} & 0.07 & 747.3k & {35.09} & \color{green}\raisebox{0ex}{$\blacktriangle$}~\textcolor{black}{\texttt{+}\hspace{0.875ex}13.98\small{\%}} & 0.19 \\
5 & 18.6k & 19.04 & \color{red}\raisebox{0ex}{$\blacktriangledown$}~\textcolor{black}{\texttt{-}\hspace{2ex}3.34\small{\%}} & 0.05 & 540.5k & {24.90} & \color{green}\raisebox{0ex}{$\blacktriangle$}~\textcolor{black}{\texttt{+}\hspace{2ex}5.22\small{\%}} & 0.09 \\
\bottomrule
        \end{tabular}
        }
    \end{minipage}
    \hfill
    \begin{minipage}{0.48\textwidth}
        \centering
        \scalebox{0.62}{
        \begin{tabular}{|c|cccr|cccr|}
            \hiderowcolors
            \toprule 
            \multicolumn{1}{c}{\raisebox{-\heavyrulewidth}{\textbf{~Dev. ID~}}} & \multicolumn{4}{c}{\textbf{Developer}} & \multicolumn{4}{c}{\textbf{Organization}}\\
            \cmidrule{1-9}
            & \textit{N°} & \textit{CB~\small\%} & \textit{$\Delta$} & \textit{$\vert ES \vert$} & \textit{N°} & \textit{CB~\small\%} & \textit{$\Delta$} & \textit{$\vert ES \vert$} \\
            \showrowcolors
            \midrule
51 & 5.6k & 19.14 & \color{white}\raisebox{0ex}{$\blacktriangle$}~\textcolor{black}{\texttt{+}\hspace{2ex}0.43\small{\%}} & 0.01 & 877.8k & {24.10} & \color{green}\raisebox{0ex}{$\blacktriangle$}~\textcolor{black}{\texttt{+}\hspace{2ex}6.03\small{\%}} & 0.09 \\
52 & 5.5k & 22.94 & \color{gray!20}\raisebox{0ex}{$\blacktriangledown$}~\textcolor{black}{\texttt{-}\hspace{2ex}0.16\small{\%}} & 0.00 & 408.3k & {25.45} & \color{gray!20}\raisebox{0ex}{$\blacktriangle$}~\textcolor{black}{\texttt{+}\hspace{2ex}1.14\small{\%}} & 0.03 \\
53 & 5.5k & 33.08 & \color{white}\raisebox{0ex}{$\blacktriangle$}~\textcolor{black}{\texttt{+}\hspace{2ex}3.11\small{\%}} & 0.06 & 846.2k & {45.48} & \color{green}\raisebox{0ex}{$\blacktriangle$}~\textcolor{black}{\texttt{+}\hspace{0.875ex}19.79\small{\%}} & 0.28 \\
54 & 5.5k & 20.19 & \color{gray!20}\raisebox{0ex}{$\blacktriangle$}~\textcolor{black}{\texttt{+}\hspace{2ex}1.49\small{\%}} & 0.02 & 211.0k & {23.60} & \color{green}\raisebox{0ex}{$\blacktriangle$}~\textcolor{black}{\texttt{+}\hspace{2ex}4.90\small{\%}} & 0.04 \\
55 & 5.3k & 18.15 & \color{white}\raisebox{0ex}{$\blacktriangle$}~\textcolor{black}{\texttt{+}\hspace{2ex}1.36\small{\%}} & 0.02 & 368.6k & {21.02} & \color{green}\raisebox{0ex}{$\blacktriangle$}~\textcolor{black}{\texttt{+}\hspace{2ex}4.01\small{\%}} & 0.06 \\
\bottomrule
        \end{tabular}
        }
    \end{minipage}
\end{sidewaystable*}
\normalsize

\begin{sidewaystable*}
    \centering
    \vspace*{15cm}
    \begin{minipage}{0.48\textwidth}
    \fontsize{12}{13}\selectfont
    \centering
    \caption{Exact Match (EM) predictions generated by the baseline and by the personalized models for Spring.}
    \label{tab:results-t5small-spring}
    \vspace*{-0.1cm}
    \rowcolors{2}{gray!20}{white}
        \centering
        \scalebox{0.62}{
        \begin{tabular}{|c|c|cccr|cccr|}
            \hiderowcolors
            \toprule 
            \multicolumn{1}{c}{\raisebox{-\heavyrulewidth}{\textbf{Dev. ID}}} & \multicolumn{1}{c}{\raisebox{-\heavyrulewidth}{\textbf{Baseline B$_{s}$}}} & \multicolumn{4}{c}{\textbf{Developer}} & \multicolumn{4}{c}{\textbf{Organization}}\\
            \cmidrule{1-10}
            & \textit{EM~\small\%} & \textit{N°} & \textit{EM~\small\%} & \textit{$\Delta$} & \textit{OR} & \textit{N°} & \textit{EM~\small\%} & \textit{$\Delta$} & \textit{OR} \\
            \showrowcolors
            \midrule
1 & 23.8 & 17.9k & 26.8 & \color{green}\raisebox{0ex}{$\blacktriangle$}~\textcolor{black}{\texttt{+}\hspace{2ex}3.00\small{\%}} & 2.67 & 228.6k & \textbf{27.4} & \color{green}\raisebox{0ex}{$\blacktriangle$}~\textcolor{black}{\texttt{+}\hspace{2ex}3.60\small{\%}} & 2.80 \\
2 & 22.8 & 16.9k & 24.4 & \color{gray!20}\raisebox{0ex}{$\blacktriangle$}~\textcolor{black}{\texttt{+}\hspace{2ex}1.60\small{\%}} & 1.53 & 188.0k & \textbf{26.0} & \color{green}\raisebox{0ex}{$\blacktriangle$}~\textcolor{black}{\texttt{+}\hspace{2ex}3.20\small{\%}} & 2.45 \\
3 & 19.6 & 13.4k & 18.2 & \color{white}\raisebox{0ex}{$\blacktriangledown$}~\textcolor{black}{\texttt{-}\hspace{2ex}1.40\small{\%}} & 0.65 & 222.0k & \textbf{21.0} & \color{white}\raisebox{0ex}{$\blacktriangle$}~\textcolor{black}{\texttt{+}\hspace{2ex}1.40\small{\%}} & 1.50 \\
4 & 18.2 & 12.7k & 19.2 & \color{gray!20}\raisebox{0ex}{$\blacktriangle$}~\textcolor{black}{\texttt{+}\hspace{2ex}1.00\small{\%}} & 1.62 & 226.5k & \textbf{19.4} & \color{gray!20}\raisebox{0ex}{$\blacktriangle$}~\textcolor{black}{\texttt{+}\hspace{2ex}1.20\small{\%}} & 1.43 \\
5 & 26.0 & 12.6k & 27.2 & \color{white}\raisebox{0ex}{$\blacktriangle$}~\textcolor{black}{\texttt{+}\hspace{2ex}1.20\small{\%}} & 1.60 & 223.8k & \textbf{28.2} & \color{white}\raisebox{0ex}{$\blacktriangle$}~\textcolor{black}{\texttt{+}\hspace{2ex}2.20\small{\%}} & 1.92 \\
\bottomrule
        \end{tabular}
        }
\end{minipage}
\hfill
\begin{minipage}{0.48\textwidth}
    \fontsize{12}{13}\selectfont
    \centering
    \caption{CrystalBLEU (CB) average score between the baseline and the personalized models for Spring.}
    \label{tab:results-t5small-spring-cb}
    \vspace*{-0.1cm}
    \rowcolors{2}{gray!20}{white}
        \centering
        \scalebox{0.62}{
        \begin{tabular}{|c|cccr|cccr|}
            \hiderowcolors
            \toprule 
            \multicolumn{1}{c}{\raisebox{-\heavyrulewidth}{\textbf{Dev. ID}}} & \multicolumn{4}{c}{\textbf{Developer}} & \multicolumn{4}{c}{\textbf{Organization}}\\
            \cmidrule{1-9}
            & \textit{N°} & \textit{CB~\small\%} & \textit{$\Delta$} & \textit{$\vert ES \vert$} & \textit{N°} & \textit{CB~\small\%} & \textit{$\Delta$} & \textit{$\vert ES \vert$} \\
            \showrowcolors
            \midrule
1 & 17.9k & 21.10 & \color{white}\raisebox{0ex}{$\blacktriangle$}~\textcolor{black}{\texttt{+}\hspace{2ex}1.98\small{\%}} & 0.02 & 228.6k & {25.13} & \color{green}\raisebox{0ex}{$\blacktriangle$}~\textcolor{black}{\texttt{+}\hspace{2ex}5.80\small{\%}} & 0.08 \\
2 & 16.9k & 22.95 & \color{green}\raisebox{0ex}{$\blacktriangle$}~\textcolor{black}{\texttt{+}\hspace{2ex}6.43\small{\%}} & 0.14 & 188.0k & {26.49} & \color{green}\raisebox{0ex}{$\blacktriangle$}~\textcolor{black}{\texttt{+}\hspace{0.875ex}10.81\small{\%}} & 0.21 \\
3 & 13.4k & 18.02 & \color{white}\raisebox{0ex}{$\blacktriangle$}~\textcolor{black}{\texttt{+}\hspace{2ex}0.15\small{\%}} & 0.02 & 222.0k & {21.63} & \color{green}\raisebox{0ex}{$\blacktriangle$}~\textcolor{black}{\texttt{+}\hspace{2ex}4.94\small{\%}} & 0.09 \\
4 & 12.7k & 17.74 & \color{green}\raisebox{0ex}{$\blacktriangle$}~\textcolor{black}{\texttt{+}\hspace{2ex}2.46\small{\%}} & 0.06 & 226.5k & {22.53} & \color{green}\raisebox{0ex}{$\blacktriangle$}~\textcolor{black}{\texttt{+}\hspace{2ex}6.05\small{\%}} & 0.14 \\
5 & 12.6k & 23.71 & \color{green}\raisebox{0ex}{$\blacktriangle$}~\textcolor{black}{\texttt{+}\hspace{2ex}4.61\small{\%}} & 0.10 & 223.8k & {27.44} & \color{green}\raisebox{0ex}{$\blacktriangle$}~\textcolor{black}{\texttt{+}\hspace{2ex}7.92\small{\%}} & 0.16 \\
\bottomrule
        \end{tabular}
        }
\end{minipage}
\end{sidewaystable*}
\normalsize

Tables~\ref{tab:results-t5small-em} and \ref{tab:results-t5small-spring} report the performance achieved by the baseline B$_{s}$ (\ie generic T5$_{small}$), the \ds and the \os models in terms of exact matches (\textit{EM \%}) on the 100 developers' test sets of Apache and the 36 developers' test sets of Spring, respectively. In addition, for the columns referring to the personalized models we also report the number of instances in the corresponding training sets (\textit{N°}), the delta in EM predictions with respect to B$_{s}$ (\textit{$\Delta$}), and the odds ratio (\textit{OR}) reported by the McNemar's test~\cite{mcnemar} (again, when tested against B$_{s}$). The symbol \color{green}{\raisebox{0ex}{$\blacktriangle$}} \color{black} associated to a \textit{$\Delta$} indicates a statistically significant increase in EM predictions with respect to B$_{s}$ ($p$-value $<$0.05); similarly, a \color{red}{\raisebox{0ex}{$\blacktriangledown$}} \color{black} indicates statistically significant decreases in performance. In each row, bold values indicate the model achieving the best performance on the corresponding test set. To make a concrete example, let us consider the results achieved on the test set related to developer 2 of Apache (\textit{Dev. ID} = 2 in \tabref{tab:results-t5small-em}): B$_{s}$ achieved 31.8\% of EM predictions on D$^{2}$'s test set, against the 33.2\% of the \ds model (trained on additional $\sim$23.2k instances), and the 37.0\% of the \os model (trained on additional $\sim$556k instances). Thus, the absolute increase in performance with respect to B$_{s}$ is +1.40\% for the \ds model (non statistically significant) and +5.20\% for the \os model. The latter increase is statistically significant, with an OR of 3.36, indicating $\sim$3 times higher odds to generate an EM prediction than B$_{s}$.

Similarly, Tables~\ref{tab:results-t5small-cb} and \ref{tab:results-t5small-spring-cb} compare the average CrystalBLEU scores (\textit{CB \%}) achieved by the B$_{s}$ baseline and by the personalized models on the developers' test sets (for Apache and Spring developers, respectively). The basic idea behind CrystalBLEU is to compare non-EM predictions generated by two models, to see whether one of the two models still generates predictions closer to the target when not outputting an EM. We exclude from this analysis cases in which both models generate an EM, since the CrystalBLEU would be trivially equal to 1 for both of them. In both tables, the gap between the average CrystalBLEU scores achieved by the personalized models and by B$_{s}$ is shown in the column \textit{$\Delta$}, while the $\vert ES \vert$ column shows the effect size (Cliff's delta \cite{Cliff:2005}) of the difference. For example, for developer 2 of Apache (\textit{Dev. ID} = 2 in \tabref{tab:results-t5small-cb}) we observe an increase of +3.63\% in the CrystalBLEU score for the \ds model as compared to B$_{s}$, and an increase of +8.68\% for the \os model. The first increase is not statistically significant, while the latter is, with an effect size of 0.13, indicating a negligible effect of the \os model in generating better predictions than B$_{s}$.

\subsection{Goal 1: Evaluating Developer-Specific Personalization}
In terms of exact match predictions (EM), 76\% (76 out of 100) of Apache \ds models benefitted from the second fine-tuning (33 out of 76 are statistically significant) with an average performance improvement of 5.37\% (median=2.00\%) and a mean OR of 3.91 (min=1.04, max=53.75). Likewise, 83\% (30 out of 36) of the Spring \ds models achieved better performance than B$_{s}$ (eight out of 30 statistically significant), with an average performance increase of 1.77\% (median=1.20\%) and a mean OR of 1.99 (min=1.17, max=6.62).
Only a small part of the \ds models did not improve performance, namely 24 models for Apache (mean=-1.23\%, median=-1.00\%) and six for Spring (mean=-0.63\%, median=-0.40\%). Out of these, only three performance decreases were statistically significant (all in Apache, \textit{Dev. ID} = 5, 40, 58). On top of these, four Apache models achieved exactly the same EM predictions of B$_{s}$ (\textit{Dev. ID} = 8, 24, 29, 32).

As it can be observed in Tables \ref{tab:results-t5small-em} and \ref{tab:results-t5small-spring}, there are several developers which clearly represent (positive) outliers in terms of achieved increase in performance (see \eg \textit{Dev. ID} = 7 for Apache and \textit{Dev. ID} = 14 for Spring). We inspected their test sets to understand why a \ds model was working so well as compared to B$_{s}$. We found that these developers implemented a significant amount of \emph{project-specific boilerplate code} (\eg \texttt{toString}, \texttt{compareTo}), thus their training and test sets shared similar code structures. These are typical code elements which nowadays can be ``delegated'' to an AI assistant. Still, the generic baseline model failed in predicting their completions due to the lack of project-specific code. As an illustration, the test set of developer 57 of Apache features a completion on the \texttt{get\-Longnvarchar\-Column} method, whose purpose is to retrieve the \texttt{longnvarchar\-Column} attribute from a data collection. The \ds model, having knowledge about the code base on which the developer works, was able to predict the need to invoke \texttt{read\-Property} to retrieve the attribute (as done in other code locations in the \ds training set), while the baseline model lacked such a piece of knowledge, recommending a wrong completion.

For what concerns the three developers on which the \ds model resulted in a statistically significant decrease of EMs, we inspected their predictions to identify patterns explaining this result. We found that sometimes these models fail in predicting the correct code (as opposed to the baseline) due to repetitions of the same suggestion multiple times or the addition of extra code tokens. Additional studies are needed to understand how to address this point, \eg adopting a smaller learning rate for the \ds fine-tuning to avoid influencing too much the model towards the developer's coding style.

Table \ref{tab:results-t5small-cb} shows that, even when focusing on non-EM predictions, 84\% of Apache developers' test sets have better predictions with the \ds model, with an average CrystalBLEU improvement of 6.63\% (median=2.52\%). Of these improvements, 34 are statistically significant, with an average (small) Cliff's delta of 0.18 (min=0.03, max=0.69). The trend observed for the Spring organization (depicted in Table~\ref{tab:results-t5small-spring-cb}) is even clearer, with all developers but one (\textit{Dev. ID} = 26) benefitting from better predictions with the \ds models (20 statistically significant, mean effect size of 0.08---min=0.03, max=0.14). The average improvement in CrystalBLEU is 2.96\% (median=2.63\%).
All this suggests that, overall, while the \ds models seem to generate predictions closer to the target even when not outputting the correct prediction (\ie overall higher CrystalBLEU), the mostly negligible/small effect size we observed tells us that the gap is not major. Still, it is worth noting that the overall positive trend observed for the \ds models (when looking at both EM predictions and CrystalBLEU) is the result of a very limited training effort, with \ds training datasets going from 1k up to 46.6k instances (column \textit{N°}). As a term of comparison, B$_{s}$ was fine-tuned on 1.4M instances.

\begin{tcolorbox}[title=\faLightbulbO~~Summary of Findings]
    The overall trend we observed is that a \ds training tends to improve the model's code completion capabilities, with improvements being statistically significant for 30\% (41 out of 136) of the studied developers in terms of EM predictions, and for 40\% of developers (54 out of 136) in terms of CrystalBLEU. For both metrics, however, the magnitude of the observed improvement is limited, but still noteworthy when considering the very limited additional training that was performed (due to the limited fine-tuning data specific of a developer).
\end{tcolorbox}

\subsection{Goal 1: Evaluating Organization-Specific Personalization}
Concerning the \os customization, for the Apache organization, 93\% of the models achieved better performance than B$_{s}$ in terms of EM predictions, with 70 of these improvements being statistically significant. The average EM improvement is +7.84\% (median=4.00\%) with a mean OR of 7.59 (min=1.05, max=143.00). Similarly, for Spring all \os models but one (\textit{Dev. ID} = 36) were better than B$_{s}$, with 17 of these improvements being statistically significant. The average EM improvement is +2.84\% (median=2.40\%) with a mean OR of 2.37 (min=1.20, max=6.00).
Only one model obtained a statistically significant decrease in performance across the two organizations (\textit{Dev. ID} = 61 of Apache).

The \os models beat the \ds ones in terms of EM predictions in 89\% of cases for Apache and in 81\% of cases for Spring. This is likely due to the larger amount of training data present in the \os training sets. The confounding factor related to the size of the training set is thoroughly discussed in \secref{sub:trainingSize}. The analysis of the CrystalBLEU supports the conclusions obtained when analyzing the EM predictions: the \os models are superior to B$_{s}$ in 96\% of cases for Apache and in every case for Spring, achieving an average CrystalBLEU improvement of 11.07\% (median=6.31) for Apache, and 6.69\% (median=6.84) for Spring. In Apache, 85\% of developers observe a statistically-significant improvement of CrystalBLEU, with an average effect size of 0.17 (min=0.04, max=0.79). In Spring, the differences are significant for 83\% of developers, with an average effect size of 0.14 (min=0.07, max=0.23). It is worth noting that, differently from what observed for the developer-specific models, the improvements seen with the organization-specific fine-tuning (i) are more consistent (\ie a higher percentage of developers benefitted from statistically-significant improvements); and (ii) are characterized by higher ORs (for EM) and effect sizes (for CrystalBLEU). 

\begin{tcolorbox}[title=\faLightbulbO~~Summary of Findings]
    The \os models are the ones providing the best performance, being the best in class for 89 (29) out of the subject 100 (36) developers. The average increase in EM predictions is +7.84\% (+2.84\%) over the baseline. The CrystalBLEU study confirms the superiority of the \os models, which outperform 96\% (100\%) of the baseline models, with an average improvement of +11.07\% (+6.69\%).
\end{tcolorbox}

\subsection{Goal 2: Assessing the Impact of the Training Data Size}
\label{sub:trainingSize}

The discussed findings indicate a ranking between the experimented models, with the \os being the best, followed by the \ds and, finally, the baseline (B$_{s}$). This ranking also reflects the amount of training data used in each training strategy (\eg \os benefited from more training data), questioning the role played by the training dataset size on the differences in performance.  

We start from the superiority demonstrated by the \os training over the \ds training. Our assumption is that the former was superior only due to the additional training data and that the latter would be superior (since it is more specific) if given the same amount of training instances. To test this assumption we train 20 additional \os models (one for each of the top-10 developers of each organization) in which we cap the size of the training data to the exact same amount of training instances we collected for the corresponding \ds model. For example, since the \ds training set for \emph{Dev. ID} = 1 of Apache has 46.6k instances, we randomly select the same number of instances from the corresponding \os dataset, training with it what we call the \emph{Organization subset} model. The left-hand side of \tabref{tab:rnd-small-results} reports the achieved results for Apache (top) and Spring (bottom). To provide more context, the table shows again the ID of the developer the test set refers to, the EM predictions achieved by the baseline (\emph{Baseline B$_{s}$} column) and by the \ds model (\emph{Developer} column) and, in addition to that, the EM predictions generated with the \emph{Organization subset} model.
As observed, when the \os models benefit from the same amount of instances as the \ds models, their performance is overall worse. Indeed,  the only statistically significant differences in terms of EM predictions (\emph{Dev. ID} = 1, 4, 7, 9 of Apache and \emph{Dev. ID} = 10 of Spring) are in favor of the \ds model and accompanied by an OR of at least 1.95. This suggests that, in the context of personalizing DL-based code completion tools, more specific data (\eg \ds data) must be preferred over more generic ones (\eg \os data) when possible.

\begin{table*}[t]
    \centering
    \caption{Comparison of Exact Match (EM) predictions with models trained on different datasets of the same size.}
    \label{tab:rnd-small-results}
    \begin{minipage}{.48\textwidth}
        \centering
        \fontsize{12}{13}\selectfont
        \vspace*{-0.1cm}
        \resizebox*{!}{0.3\textheight}{
        \rowcolors{2}{gray!20}{white}
        \begin{tabular}{c|c|c|cc|ccr}
            \hiderowcolors
            \toprule 
            & \multicolumn{1}{c}{\raisebox{-\heavyrulewidth}{\textbf{Dev. ID}}} & \multicolumn{1}{c}{\raisebox{-\heavyrulewidth}{\textbf{Baseline B$_{s}$}}} & \multicolumn{2}{c}{\textbf{Developer}} & \multicolumn{3}{c}{\textbf{Organization subset}}\\
            \cmidrule{2-8}
            & & \textit{EM~\small\%} & \textit{N°} & \textit{EM~\small\%} & \textit{EM~\small\%} & \textit{$\Delta$} & \textit{OR} \\
            \midrule
            \multirow{10}{*}{\rotatebox[origin=c]{90}{Apache}} & 1 & 51.2 & 46.6k & 60.8 & 57.2 & \color{red}\raisebox{0ex}{$\blacktriangledown$}~\textcolor{black}{\texttt{-}\hspace{2ex}3.60\small{\%}} & 2.38 \\
& 2 & 31.8 & 23.2k & 33.2 & 33.4 & \color{white}\raisebox{0ex}{$\blacktriangle$}~\textcolor{black}{\texttt{+}\hspace{2ex}0.20\small{\%}} & 0.96 \\
& 3 & 26.0 & 20.3k & 25.0 & 24.8 & \color{white}\raisebox{0ex}{$\blacktriangledown$}~\textcolor{black}{\texttt{-}\hspace{2ex}0.20\small{\%}} & 1.07 \\
& 4 & 35.2 & 19.2k & 39.0 & 35.2 & \color{red}\raisebox{0ex}{$\blacktriangledown$}~\textcolor{black}{\texttt{-}\hspace{2ex}3.80\small{\%}} & 1.95 \\
& 5 & 21.4 & 18.6k & 18.6 & 20.4 & \color{white}\raisebox{0ex}{$\blacktriangle$}~\textcolor{black}{\texttt{+}\hspace{2ex}1.80\small{\%}} & 0.50 \\
& 6 & 34.6 & 17.8k & 36.8 & 36.2 & \color{white}\raisebox{0ex}{$\blacktriangledown$}~\textcolor{black}{\texttt{-}\hspace{2ex}0.60\small{\%}} & 1.21 \\
& 7 & 28.8 & 17.5k & 60.0 & 51.8 & \color{red}\raisebox{0ex}{$\blacktriangledown$}~\textcolor{black}{\texttt{-}\hspace{2ex}8.20\small{\%}} & 2.64 \\
& 8 & 29.0 & 17.0k & 29.0 & 28.4 & \color{white}\raisebox{0ex}{$\blacktriangledown$}~\textcolor{black}{\texttt{-}\hspace{2ex}0.60\small{\%}} & 1.30 \\
& 9 & 27.8 & 15.9k & 46.2 & 37.2 & \color{red}\raisebox{0ex}{$\blacktriangledown$}~\textcolor{black}{\texttt{-}\hspace{2ex}9.00\small{\%}} & 3.50 \\
& 10 & 31.6 & 15.8k & 33.2 & 33.2 & \color{white}\raisebox{0.5ex}{\rule{0.60em}{0.2em}}~\textcolor{black}{\texttt{ }\hspace{2ex}0.00\small{\%}} & 1.00 \\
\bottomrule
\multirow{10}{*}{\rotatebox[origin=c]{90}{Spring}}  & 1 & 23.8 & 17.9k & 26.8 & 24.8 & \color{white}\raisebox{0ex}{$\blacktriangledown$}~\textcolor{black}{\texttt{-}\hspace{2ex}2.00\small{\%}} & 2.11 \\
& 2 & 22.8 & 16.9k & 24.4 & 22.2 & \color{white}\raisebox{0ex}{$\blacktriangledown$}~\textcolor{black}{\texttt{-}\hspace{2ex}2.20\small{\%}} & 2.00 \\
& 3 & 19.6 & 13.4k & 18.2 & 19.0 & \color{white}\raisebox{0ex}{$\blacktriangle$}~\textcolor{black}{\texttt{+}\hspace{2ex}0.80\small{\%}} & 0.69 \\
& 4 & 18.2 & 12.7k & 19.2 & 17.8 & \color{white}\raisebox{0ex}{$\blacktriangledown$}~\textcolor{black}{\texttt{-}\hspace{2ex}1.40\small{\%}} & 2.00 \\
& 5 & 26.0 & 12.6k & 27.2 & 27.2 & \color{white}\raisebox{0.5ex}{\rule{0.60em}{0.2em}}~\textcolor{black}{\texttt{ }\hspace{2ex}0.00\small{\%}} & 1.00 \\
& 6 & 21.0 & 11.4k & 22.2 & 21.0 & \color{white}\raisebox{0ex}{$\blacktriangledown$}~\textcolor{black}{\texttt{-}\hspace{2ex}1.20\small{\%}} & 1.67 \\
& 7 & 24.4 & 11.2k & 25.4 & 27.0 & \color{white}\raisebox{0ex}{$\blacktriangle$}~\textcolor{black}{\texttt{+}\hspace{2ex}1.60\small{\%}} & 0.50 \\
& 8 & 18.2 & 10.7k & 21.0 & 19.6 & \color{white}\raisebox{0ex}{$\blacktriangledown$}~\textcolor{black}{\texttt{-}\hspace{2ex}1.40\small{\%}} & 1.88 \\
& 9 & 22.6 & 10.3k & 23.6 & 24.0 & \color{white}\raisebox{0ex}{$\blacktriangle$}~\textcolor{black}{\texttt{+}\hspace{2ex}0.40\small{\%}} & 0.91 \\
& 10 & 21.4 & 10.1k & 25.8 & 22.0 & \color{red}\raisebox{0ex}{$\blacktriangledown$}~\textcolor{black}{\texttt{-}\hspace{2ex}3.80\small{\%}} & 3.71 \\
\bottomrule
 \end{tabular}
        }
    \end{minipage}%
    \begin{minipage}{.48\textwidth}
        \centering
        \fontsize{12}{13}\selectfont
        \vspace*{-0.1cm}
        \resizebox*{!}{0.3\textheight}{
        \rowcolors{2}{gray!20}{white}
        \begin{tabular}{c|c|c|cc|ccr}
            \hiderowcolors
            \toprule 
            & \multicolumn{1}{c}{\raisebox{-\heavyrulewidth}{\textbf{Dev. ID}}} & \multicolumn{1}{c}{\raisebox{-\heavyrulewidth}{\textbf{Baseline B$_{s}$}}} & \multicolumn{2}{c}{\textbf{Organization}} & \multicolumn{3}{c}{\textbf{Baseline+}}\\
            \cmidrule{2-8}
            & & \textit{EM~\small\%} & \textit{N°} & \textit{EM~\small\%} & \textit{EM~\small\%} & \textit{$\Delta$} & \textit{OR} \\
            \midrule
            \multirow{10}{*}{\rotatebox[origin=c]{90}{Apache}} & 1 & 51.2 & 888.0k & 61.8 & 48.8 & \color{red}\raisebox{0ex}{$\blacktriangledown$}~\textcolor{black}{\texttt{-}\hspace{0.875ex}13.00\small{\%}} & 9.12 \\
& 2 & 31.8 & 556.0k & 37.0 & 32.6 & \color{red}\raisebox{0ex}{$\blacktriangledown$}~\textcolor{black}{\texttt{-}\hspace{2ex}4.40\small{\%}} & 3.00 \\
& 3 & 26.0 & 830.8k & 28.2 & 27.2 & \color{white}\raisebox{0ex}{$\blacktriangledown$}~\textcolor{black}{\texttt{-}\hspace{2ex}1.00\small{\%}} & 1.42 \\
& 4 & 35.2 & 747.3k & 44.2 & 39.8 & \color{red}\raisebox{0ex}{$\blacktriangledown$}~\textcolor{black}{\texttt{-}\hspace{2ex}4.40\small{\%}} & 3.00 \\
& 5 & 21.4 & 540.5k & 24.0 & 24.4 & \color{white}\raisebox{0ex}{$\blacktriangle$}~\textcolor{black}{\texttt{+}\hspace{2ex}0.40\small{\%}} & 0.89 \\
& 6 & 34.6 & 791.4k & 37.2 & 38.4 & \color{white}\raisebox{0ex}{$\blacktriangle$}~\textcolor{black}{\texttt{+}\hspace{2ex}1.20\small{\%}} & 0.73 \\
& 7 & 28.8 & 524.8k & 70.0 & 46.4 & \color{red}\raisebox{0ex}{$\blacktriangledown$}~\textcolor{black}{\texttt{-}\hspace{0.875ex}23.60\small{\%}} & 14.11 \\
& 8 & 29.0 & 850.5k & 32.2 & 30.4 & \color{white}\raisebox{0ex}{$\blacktriangledown$}~\textcolor{black}{\texttt{-}\hspace{2ex}1.80\small{\%}} & 1.60 \\
& 9 & 27.8 & 580.9k & 48.4 & 38.0 & \color{red}\raisebox{0ex}{$\blacktriangledown$}~\textcolor{black}{\texttt{-}\hspace{0.875ex}10.40\small{\%}} & 5.33 \\
& 10 & 31.6 & 146.3k & 33.6 & 33.8 & \color{white}\raisebox{0ex}{$\blacktriangle$}~\textcolor{black}{\texttt{+}\hspace{2ex}0.20\small{\%}} & 0.96 \\
\bottomrule
\multirow{10}{*}{\rotatebox[origin=c]{90}{Spring}} & 1 & 23.8 & 228.6k & 27.4 & 24.8 & \color{red}\raisebox{0ex}{$\blacktriangledown$}~\textcolor{black}{\texttt{-}\hspace{2ex}2.60\small{\%}} & 2.18 \\
& 2 & 22.8 & 188.0k & 26.0 & 22.0 & \color{red}\raisebox{0ex}{$\blacktriangledown$}~\textcolor{black}{\texttt{-}\hspace{2ex}4.00\small{\%}} & 3.00 \\
& 3 & 19.6 & 222.0k & 21.0 & 17.8 & \color{red}\raisebox{0ex}{$\blacktriangledown$}~\textcolor{black}{\texttt{-}\hspace{2ex}3.20\small{\%}} & 2.45 \\
& 4 & 18.2 & 226.5k & 19.4 & 17.4 & \color{white}\raisebox{0ex}{$\blacktriangledown$}~\textcolor{black}{\texttt{-}\hspace{2ex}2.00\small{\%}} & 1.91 \\
& 5 & 26.0 & 223.8k & 28.2 & 28.2 & \color{white}\raisebox{0.5ex}{\rule{0.60em}{0.2em}}~\textcolor{black}{\texttt{ }\hspace{2ex}0.00\small{\%}} & 1.00 \\
& 6 & 21.0 & 243.1k & 25.6 & 20.4 & \color{red}\raisebox{0ex}{$\blacktriangledown$}~\textcolor{black}{\texttt{-}\hspace{2ex}5.20\small{\%}} & 4.25 \\
& 7 & 24.4 & 201.5k & 27.8 & 25.4 & \color{white}\raisebox{0ex}{$\blacktriangledown$}~\textcolor{black}{\texttt{-}\hspace{2ex}2.40\small{\%}} & 2.20 \\
& 8 & 18.2 & 193.0k & 22.2 & 19.4 & \color{red}\raisebox{0ex}{$\blacktriangledown$}~\textcolor{black}{\texttt{-}\hspace{2ex}2.80\small{\%}} & 3.80 \\
& 9 & 22.6 & 194.7k & 26.4 & 22.6 & \color{red}\raisebox{0ex}{$\blacktriangledown$}~\textcolor{black}{\texttt{-}\hspace{2ex}3.80\small{\%}} & 2.58 \\
& 10 & 21.4 & 230.8k & 23.8 & 20.6 & \color{red}\raisebox{0ex}{$\blacktriangledown$}~\textcolor{black}{\texttt{-}\hspace{2ex}3.20\small{\%}} & 2.07 \\
\bottomrule
\end{tabular}
}
\end{minipage}
\end{table*} 
\normalsize
\begin{table*}[t]
    \centering
    \caption{Comparison of CrystalBLEU (CB) scores with models trained on different datasets of the same size.}
    \label{tab:rnd-small-results-cb}
    \begin{minipage}{.49\textwidth}
        \centering
        \fontsize{12}{13}\selectfont
        \vspace*{-0.1cm}
        \resizebox*{!}{0.35\textheight}{
        \rowcolors{2}{gray!20}{white}
        \begin{tabular}{c|c|cc|ccr}
            \hiderowcolors
            \toprule 
            & \multicolumn{1}{c}{\raisebox{-\heavyrulewidth}{\textbf{~Dev. ID~}}} & \multicolumn{2}{c}{\textbf{Developer}} & \multicolumn{3}{c}{\textbf{Organization subset}}\\
            \cmidrule{2-7}
             & & \textit{N°} & \textit{CB~\small\%} & \textit{CB~\small\%} & \textit{$\Delta$} & \textit{$\vert ES \vert$} \\
            \midrule
            \multirow{10}{*}{\rotatebox[origin=c]{90}{Apache}} & 1 & 46.6k & 31.17 & 23.47 & \color{red}\raisebox{0ex}{$\blacktriangledown$}~\textcolor{black}{\texttt{-}\hspace{2ex}7.70\small{\%}} & 0.10 \\
& 2 & 23.2k & 27.36 & 24.97 & \color{white}\raisebox{0ex}{$\blacktriangledown$}~\textcolor{black}{\texttt{-}\hspace{2ex}2.39\small{\%}} & 0.04 \\
& 3 & 20.3k & 21.26 & 20.76 & \color{white}\raisebox{0ex}{$\blacktriangledown$}~\textcolor{black}{\texttt{-}\hspace{2ex}0.50\small{\%}} & 0.02 \\
& 4 & 19.2k & 27.44 & 22.32 & \color{red}\raisebox{0ex}{$\blacktriangledown$}~\textcolor{black}{\texttt{-}\hspace{2ex}5.12\small{\%}} & 0.06 \\
& 5 & 18.6k & 18.84 & 20.64 & \color{green}\raisebox{0ex}{$\blacktriangle$}~\textcolor{black}{\texttt{+}\hspace{2ex}1.80\small{\%}} & 0.02 \\
& 6 & 17.8k & 22.20 & 18.30 & \color{red}\raisebox{0ex}{$\blacktriangledown$}~\textcolor{black}{\texttt{-}\hspace{2ex}3.90\small{\%}} & 0.06 \\
& 7 & 17.5k & 39.92 & 33.10 & \color{red}\raisebox{0ex}{$\blacktriangledown$}~\textcolor{black}{\texttt{-}\hspace{2ex}6.82\small{\%}} & 0.08 \\
& 8 & 17.0k & 18.82 & 19.36 & \color{white}\raisebox{0ex}{$\blacktriangle$}~\textcolor{black}{\texttt{+}\hspace{2ex}0.54\small{\%}} & 0.02 \\
& 9 & 15.9k & 34.75 & 21.91 & \color{red}\raisebox{0ex}{$\blacktriangledown$}~\textcolor{black}{\texttt{-}\hspace{0.875ex}12.84\small{\%}} & 0.16 \\
& 10 & 15.8k & 29.30 & 28.16 & \color{white}\raisebox{0ex}{$\blacktriangledown$}~\textcolor{black}{\texttt{-}\hspace{2ex}1.14\small{\%}} & 0.03 \\
\bottomrule
\multirow{10}{*}{\rotatebox[origin=c]{90}{Spring}} & 1 & 17.9k & 20.08 & 18.81 & \color{white}\raisebox{0ex}{$\blacktriangledown$}~\textcolor{black}{\texttt{-}\hspace{2ex}1.27\small{\%}} & 0.00 \\
& 2 & 16.9k & 22.75 & 18.62 & \color{red}\raisebox{0ex}{$\blacktriangledown$}~\textcolor{black}{\texttt{-}\hspace{2ex}4.13\small{\%}} & 0.06 \\
& 3 & 13.4k & 17.24 & 17.60 & \color{white}\raisebox{0ex}{$\blacktriangle$}~\textcolor{black}{\texttt{+}\hspace{2ex}0.36\small{\%}} & 0.00 \\
& 4 & 12.7k & 17.93 & 15.66 & \color{red}\raisebox{0ex}{$\blacktriangledown$}~\textcolor{black}{\texttt{-}\hspace{2ex}2.27\small{\%}} & 0.03 \\
& 5 & 12.6k & 22.90 & 21.54 & \color{white}\raisebox{0ex}{$\blacktriangledown$}~\textcolor{black}{\texttt{-}\hspace{2ex}1.36\small{\%}} & 0.04 \\
& 6 & 11.4k & 20.01 & 17.06 & \color{red}\raisebox{0ex}{$\blacktriangledown$}~\textcolor{black}{\texttt{-}\hspace{2ex}2.95\small{\%}} & 0.08 \\
& 7 & 11.2k & 17.59 & 18.73 & \color{white}\raisebox{0ex}{$\blacktriangle$}~\textcolor{black}{\texttt{+}\hspace{2ex}1.14\small{\%}} & 0.01 \\
& 8 & 10.7k & 18.70 & 16.63 & \color{red}\raisebox{0ex}{$\blacktriangledown$}~\textcolor{black}{\texttt{-}\hspace{2ex}2.07\small{\%}} & 0.03 \\
& 9 & 10.3k & 23.99 & 24.32 & \color{white}\raisebox{0ex}{$\blacktriangle$}~\textcolor{black}{\texttt{+}\hspace{2ex}0.33\small{\%}} & 0.00 \\
& 10 & 10.1k & 19.49 & 16.28 & \color{white}\raisebox{0ex}{$\blacktriangledown$}~\textcolor{black}{\texttt{-}\hspace{2ex}3.21\small{\%}} & 0.03 \\
\bottomrule
 \end{tabular}
        }
    \end{minipage}%
    \begin{minipage}{.49\textwidth}
        \centering
        \fontsize{12}{13}\selectfont
        \vspace*{-0.1cm}
        \resizebox*{!}{0.35\textheight}{
        \rowcolors{2}{gray!20}{white}
        \begin{tabular}{c|c|cc|ccr}
            \hiderowcolors
            \toprule 
            & \multicolumn{1}{c}{\raisebox{-\heavyrulewidth}{\textbf{~Dev. ID~}}} & \multicolumn{2}{c}{\textbf{Organization}} & \multicolumn{3}{c}{\textbf{Baseline+}}\\
            \cmidrule{2-7}
            & & \textit{N°} & \textit{CB~\small\%} & \textit{CB~\small\%} & \textit{$\Delta$} & \textit{$\vert ES \vert$} \\
            \midrule
            \multirow{10}{*}{\rotatebox[origin=c]{90}{Apache}}&  1 & 888.0k & 44.17 & 20.83 & \color{red}\raisebox{0ex}{$\blacktriangledown$}~\textcolor{black}{\texttt{-}\hspace{0.875ex}23.34\small{\%}} & 0.29 \\
& 2 & 556.0k & 30.03 & 23.58 & \color{red}\raisebox{0ex}{$\blacktriangledown$}~\textcolor{black}{\texttt{-}\hspace{2ex}6.45\small{\%}} & 0.10 \\
& 3 & 830.8k & 25.33 & 25.19 & \color{white}\raisebox{0ex}{$\blacktriangledown$}~\textcolor{black}{\texttt{-}\hspace{2ex}0.14\small{\%}} & 0.00 \\
& 4 & 747.3k & 29.05 & 22.24 & \color{red}\raisebox{0ex}{$\blacktriangledown$}~\textcolor{black}{\texttt{-}\hspace{2ex}6.81\small{\%}} & 0.10 \\
& 5 & 540.5k & 23.77 & 24.80 & \color{white}\raisebox{0ex}{$\blacktriangle$}~\textcolor{black}{\texttt{+}\hspace{2ex}1.03\small{\%}} & 0.02 \\
& 6 & 791.4k & 22.86 & 25.08 & \color{white}\raisebox{0ex}{$\blacktriangle$}~\textcolor{black}{\texttt{+}\hspace{2ex}2.22\small{\%}} & 0.05 \\
& 7 & 524.8k & 59.62 & 31.03 & \color{red}\raisebox{0ex}{$\blacktriangledown$}~\textcolor{black}{\texttt{-}\hspace{0.875ex}28.59\small{\%}} & 0.40 \\
& 8 & 850.5k & 24.23 & 23.04 & \color{white}\raisebox{0ex}{$\blacktriangledown$}~\textcolor{black}{\texttt{-}\hspace{2ex}1.19\small{\%}} & 0.01 \\
& 9 & 580.9k & 39.55 & 24.19 & \color{red}\raisebox{0ex}{$\blacktriangledown$}~\textcolor{black}{\texttt{-}\hspace{0.875ex}15.36\small{\%}} & 0.22 \\
& 10 & 146.3k & 28.77 & 30.01 & \color{white}\raisebox{0ex}{$\blacktriangle$}~\textcolor{black}{\texttt{+}\hspace{2ex}1.24\small{\%}} & 0.01 \\
\bottomrule
\multirow{10}{*}{\rotatebox[origin=c]{90}{Spring}} &1 & 228.6k & 24.35 & 17.43 & \color{red}\raisebox{0ex}{$\blacktriangledown$}~\textcolor{black}{\texttt{-}\hspace{2ex}6.92\small{\%}} & 0.11 \\
& 2 & 188.0k & 27.04 & 17.20 & \color{red}\raisebox{0ex}{$\blacktriangledown$}~\textcolor{black}{\texttt{-}\hspace{2ex}9.84\small{\%}} & 0.20 \\
& 3 & 222.0k & 22.74 & 17.80 & \color{red}\raisebox{0ex}{$\blacktriangledown$}~\textcolor{black}{\texttt{-}\hspace{2ex}4.94\small{\%}} & 0.08 \\
& 4 & 226.5k & 22.72 & 16.37 & \color{red}\raisebox{0ex}{$\blacktriangledown$}~\textcolor{black}{\texttt{-}\hspace{2ex}6.35\small{\%}} & 0.14 \\
& 5 & 223.8k & 26.28 & 21.74 & \color{red}\raisebox{0ex}{$\blacktriangledown$}~\textcolor{black}{\texttt{-}\hspace{2ex}4.54\small{\%}} & 0.11 \\
& 6 & 243.1k & 27.73 & 15.67 & \color{red}\raisebox{0ex}{$\blacktriangledown$}~\textcolor{black}{\texttt{-}\hspace{0.875ex}12.06\small{\%}} & 0.24 \\
& 7 & 201.5k & 22.61 & 15.81 & \color{red}\raisebox{0ex}{$\blacktriangledown$}~\textcolor{black}{\texttt{-}\hspace{2ex}6.80\small{\%}} & 0.14 \\
& 8 & 193.0k & 24.51 & 15.26 & \color{red}\raisebox{0ex}{$\blacktriangledown$}~\textcolor{black}{\texttt{-}\hspace{2ex}9.25\small{\%}} & 0.22 \\
& 9 & 194.7k & 28.68 & 22.50 & \color{red}\raisebox{0ex}{$\blacktriangledown$}~\textcolor{black}{\texttt{-}\hspace{2ex}6.18\small{\%}} & 0.10 \\
& 10 & 230.8k & 25.53 & 19.58 & \color{red}\raisebox{0ex}{$\blacktriangledown$}~\textcolor{black}{\texttt{-}\hspace{2ex}5.95\small{\%}} & 0.10 \\
\bottomrule
\end{tabular}
}
\end{minipage}
\end{table*} 
\normalsize

The second comparison we perform (right-hand side of \tabref{tab:rnd-small-results}) concerns the \os models (\ie the best in class in our study) against a baseline (\emph{Baseline+} in the table) further trained on the same amount of training instances provided to the \os models, although not specific to the organization (see \secref{sec:controlling-size} for details). When comparing \emph{Baseline+} to the \os models, for 60\% of developers (\emph{Dev. ID} = 1, 2, 4, 7, 9 of Apache and \emph{Dev. ID} = 1, 2, 3, 6, 8, 9, 10 of Spring) the \os models achieve statistically significant better results in terms of EM, with an OR of at least 2.07. In no cases \emph{Baseline+} achieves statistically significant better results. Again, this supports the idea that training on more specific data (in this case, \os \emph{vs} general) helps in boosting performance.

The analysis of the CrystalBLEU, available in \tabref{tab:rnd-small-results-cb}, confirms our findings in terms of EM predictions. Predictions of \ds models have higher similarity than \emph{Organization subset} models for eight developers out of 10 for Apache (five statistically significant) and seven out of 10 for Spring (four statistically significant), while \os models outperform the \emph{Baseline+} models in seven cases out of 10 for Apache (five of which are statistically significant) and in all cases (all statistically significant) for Spring. Only developer 5 of Apache shows a statistically significant increase in favor of the \emph{Organization subset} model.

\begin{tcolorbox}[title=\faLightbulbO~~Summary of Findings]
    The amount of training instances, as expected, plays a role. Indeed, \os models are better than \ds only when the former exploit more training data than the latter (otherwise, the opposite is true). However, when controlling for the training size (and, thus, for the training cost), our findings suggest that the more specific the training data, the higher the boost in performance at inference time.
\end{tcolorbox}

\begin{figure*}[h]
\centering
\subfloat[Percentage of instances in the test set whose method signature is also present in the training set.]{\includegraphics[width=0.32\textwidth]{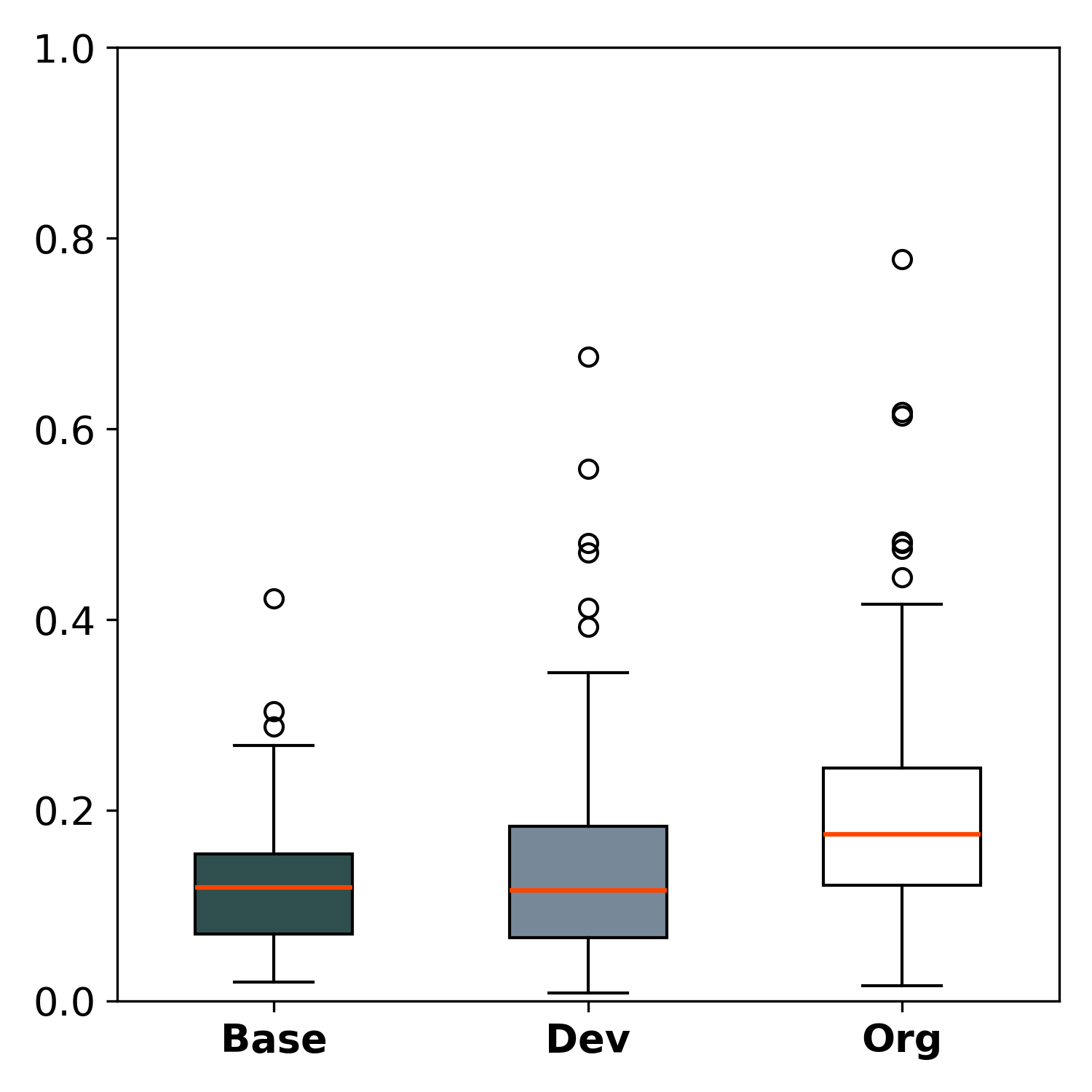}\label{fig:boxplot_signature}}
\hfill
\subfloat[Percentage of identifiers and literals in the test set which are also present in the training set.]{\includegraphics[width=0.32\textwidth]{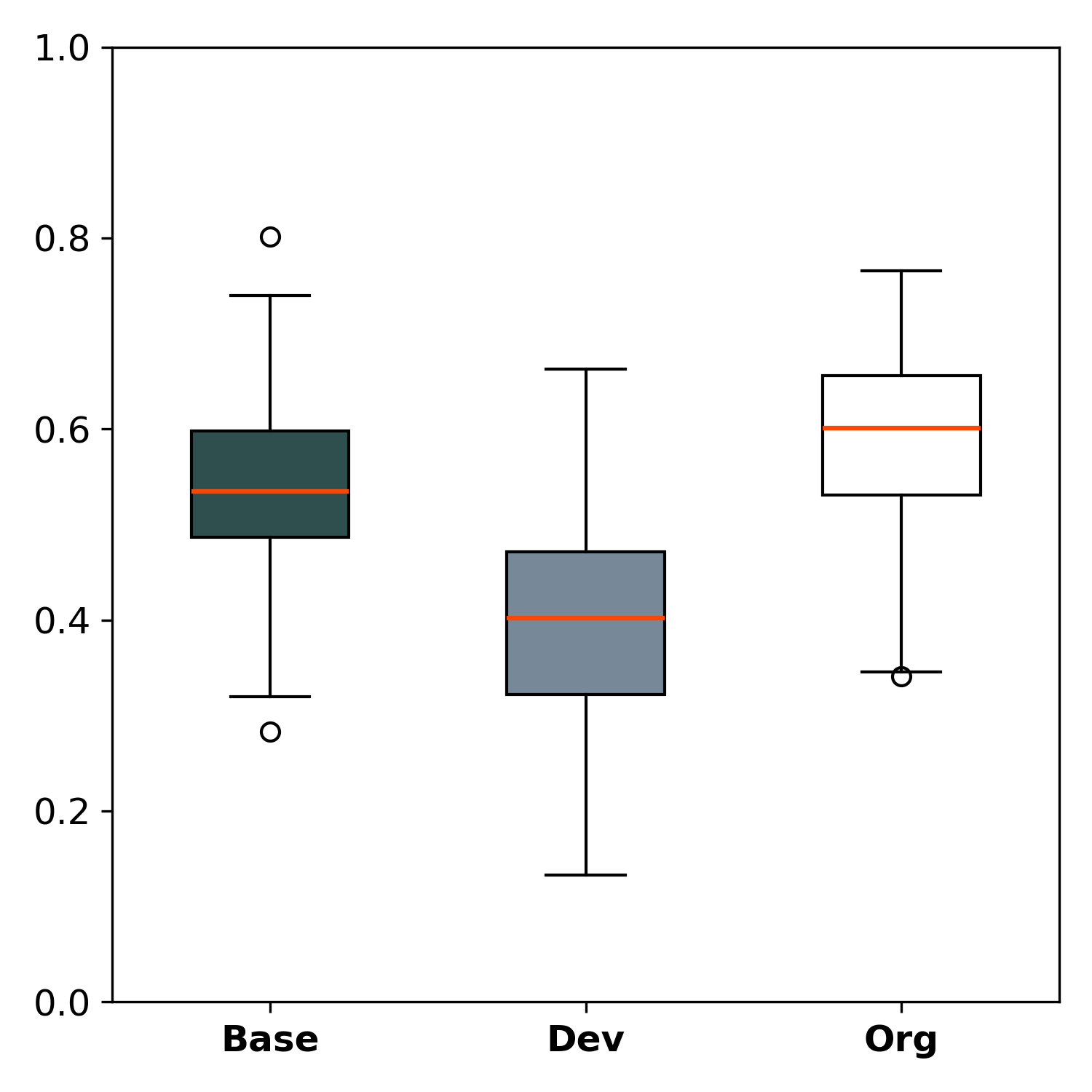}\label{fig:boxplot_id_lit_test}}
\hfill
\subfloat[Percentage of identifiers and literals in the train set which are also present in the test set.]{\includegraphics[width=0.32\textwidth]{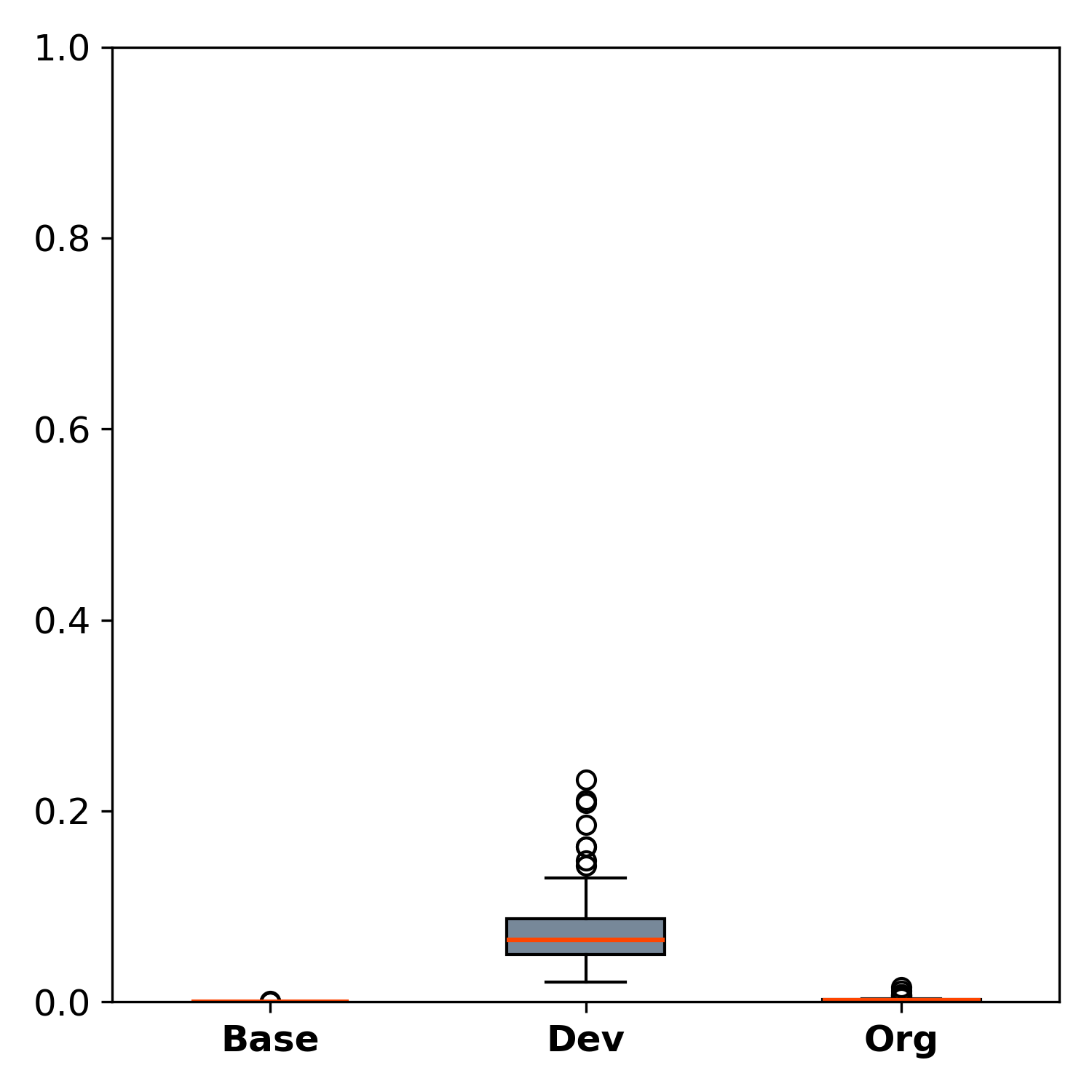}\label{fig:boxplot_id_lit_train}}

\caption{Distributions of metrics correlating training and test sets across all 100 developers of Apache. Base = Baseline B${_s}$, Dev = Developer, Org = Organization. Higher is better.}\label{fig:boxplots}
\vspace{-0.3cm}
\end{figure*}

\subsubsection{Why Do More Specific Training Data Help?}
\figref{fig:boxplots} illustrates boxplots showing \emph{information items} shared between the three trainings (Base = baseline B$_s$, Dev = \ds, Org = \os) and the 136 developer-related test sets of the two organizations, as described in \secref{sec:specific-data}.

\figref{fig:boxplot_signature} depicts the percentage of instances in the test sets whose method signature appears in the training sets.
Despite being substantially smaller, the \ds training sets cover a similar number of signatures (median=0.116) than the baseline (median=0.119), while the \os training sets cover the most signatures (median=0.175).

\figref{fig:boxplot_id_lit_test} depicts the percentage of literals (\eg strings and numbers) and identifiers (\eg method and variable/constant names) in the test sets which are also present in the training sets. As observed, the \os training sets have the highest vocabulary coverage (median=0.58), followed by the baseline (median=0.53), and the \ds (median=0.38).
Despite the lower vocabulary coverage of the \ds training sets, these are more aligned to the vocabulary used in the test sets, as illustrated in \figref{fig:boxplot_id_lit_train}, explained next.

\figref{fig:boxplot_id_lit_train} shows the percentage of identifiers and literals in the training sets which can be found in the test sets. As expected, the \ds training sets hold a larger ratio of relevant data (median=0.07) compared to the \os and baseline training sets, whose proportion is negligible (median=$\sim$0).

\begin{tcolorbox}[title=\faLightbulbO~~Summary of Findings]
    More specific training data helps in better aligning the domain of the model to the one of the test set (\figref{fig:boxplot_signature}). Also, the model will be more focused on the vocabulary used in the test set (Figures \ref{fig:boxplot_id_lit_test} and \ref{fig:boxplot_id_lit_train}).
\end{tcolorbox}

\subsection{Goal 3: Evaluating the Impact of Model Size, Architecture, and Pre-training}
In this section we analyze the extent to which our findings generalize to larger models, different architectures, and pre-trained models whose training data might already include the organizations of interest (Apache and Spring).

We start by analyzing the generalizability to larger models compared to the T5$_{small}$ (60M parameters) subject of the previous analyses. We replicate our experiments with our B$_{l}$ baseline, \ie T5$_{large}$, which features 750M parameters (12.5 times more than T5$_{small}$). To make the experimentation affordable, the experiments are replicated only for the top-10 developers from each organization (\ie 20 developers in total). \tabref{tab:results-t5large-em} reports the results achieved for Apache (top) and Spring (bottom) developers. As illustrated, the main findings previously discussed for T5$_{small}$ apply to the experiments with T5$_{large}$ as well: (i)~the more specific the training data, the higher the boost in performance; and (ii)~the \os models confirm their superiority due the larger amount of training data. Indeed, in terms of exact matches, we observe higher accuracy for \textit{developer-} and \textit{organization-specific} models with respect to B$_{l}$. Furthermore, the \textit{organization-specific} personalization reports a statistically significant increase in performance for six and eight out of 10 developers for Apache and Spring, respectively, with an average OR of 4.05 (min=2.00, max=12.33). This trend in terms of EM predictions is also confirmed by the analysis of the CrystalBLEU score. \tabref{tab:results-t5large-cb} reports the increase in the CrystalBLEU of the \ds and \os models with respect to the baseline B$_{l}$. As shown, such increase is highest for the \os models, being statistically significant for 95\% of them (19 out of 20), with an average effect size of 0.18 (min=0.08, max=0.48).

In addition to the experiments with T5$_{large}$, we also investigate the applicability of personalization to larger pre-trained code models. We focus on Code Llama in its base version featuring 7B parameters. This allows to understand whether our findings generalize also to even larger models (10 times bigger than T5$_{large}$), with different architectures (Llama-based) and already pre-trained on data likely to also feature code from the organizations of interest (since its training cutoff date is 2023~\cite{roziere2023code}). Indeed, even if Code Llama training set includes code from Apache or Spring, it is still worthwhile investigating whether further specialization can boost its performance.

\vspace{-0.5cm}
\noindent\begin{minipage}[t]{0.30\textwidth}
    \lstset{language=Java, caption={Masked method.}, label={lst:masked-code}}
    \begin{lstlisting}
public int add(int a, int b) {
    result = (*@\pink{<FILL\_ME>}@*)
    return result;
}
\end{lstlisting}
\end{minipage}
\hfill
\begin{minipage}[t]{0.36\textwidth}
    \lstset{language=Java, caption={Output by Code Llama.}, label={lst:codellama-output}}
    \begin{lstlisting}
public int add(int a, int b) {
    result = (*@\pink{a + b;}@*)
(*@\pink{\hspace{0.56cm}return result;}@*)
(*@\pink{\}}@*)
(*@\pink{\textbf{public int} subtract(\textbf{int} a, \textbf{int} b) \{}@*)
(*@\pink{\hspace{0.56cm}result = a - b;}@*)
    return result;
}
\end{lstlisting}
\end{minipage}
\hfill
\begin{minipage}[t]{0.30\textwidth}
    \lstset{language=Java, caption={Expected output.}, label={lst:expected-output}}
    \begin{lstlisting}
public int add(int a, int b) {
    result = (*@\pink{a + b;}@*)
    return result;
}
\end{lstlisting}
\end{minipage}

\begin{table*}[htpb] 
    \fontsize{12}{13}\selectfont
    \centering
    \caption{Exact Match (EM) predictions generated by the baseline and by the personalized models using T5$_{large}$.}\vspace{-0.2cm}
    \label{tab:results-t5large-em}
    \vspace*{-0.1cm}
    \resizebox*{!}{0.4\textheight}{
    \rowcolors{2}{gray!20}{white}
    \begin{tabular}{c|c|c|cccr|cccr}
        \hiderowcolors
        \toprule 
        & \multicolumn{1}{c}{\raisebox{-\heavyrulewidth}{\textbf{~Dev. ID~}}} & \multicolumn{1}{c}{\raisebox{-\heavyrulewidth}{\textbf{~Baseline B$_{l}$~}}} & \multicolumn{4}{c}{\textbf{Developer}} & \multicolumn{4}{c}{\textbf{Organization}} \\
        \cmidrule{1-11}
        & & \textit{EM~\small\%} & \textit{N°} & \textit{EM~\small\%} & \textit{$\Delta$} & \textit{OR} & \textit{N°} & \textit{EM~\small\%} & \textit{$\Delta$} & \textit{OR} \\
        \midrule
        \multirow{10}{*}{\rotatebox[origin=c]{90}{Apache}} & 1 & 52.8 & 46.6k & 64.2 & \color{green}\raisebox{0ex}{$\blacktriangle$}~\textcolor{black}{\texttt{+}\hspace{0.875ex}11.40\small{\%}} & 10.50 & 888.0k & \textbf{66.2} & \color{green}\raisebox{0ex}{$\blacktriangle$}~\textcolor{black}{\texttt{+}\hspace{0.875ex}13.40\small{\%}} & 5.79 \\
& 2 & 37.4 & 23.2k & 39.8 & \color{white}\raisebox{0ex}{$\blacktriangle$}~\textcolor{black}{\texttt{+}\hspace{2ex}2.40\small{\%}} & 1.48 & 556.0k & \textbf{44.4} & \color{green}\raisebox{0ex}{$\blacktriangle$}~\textcolor{black}{\texttt{+}\hspace{2ex}7.00\small{\%}} & 2.94 \\
& 3 & 30.6 & 20.3k & 29.4 & \color{white}\raisebox{0ex}{$\blacktriangledown$}~\textcolor{black}{\texttt{-}\hspace{2ex}1.20\small{\%}} & 0.68 & 830.8k & \textbf{33.0} & \color{white}\raisebox{0ex}{$\blacktriangle$}~\textcolor{black}{\texttt{+}\hspace{2ex}2.40\small{\%}} & 1.80 \\
& 4 & 42.6 & 19.2k & \textbf{45.4} & \color{green}\raisebox{0ex}{$\blacktriangle$}~\textcolor{black}{\texttt{+}\hspace{2ex}2.80\small{\%}} & 2.40 & 747.3k & 44.4 & \color{white}\raisebox{0ex}{$\blacktriangle$}~\textcolor{black}{\texttt{+}\hspace{2ex}1.80\small{\%}} & 1.56 \\
& 5 & 29.0 & 18.6k & 29.2 & \color{white}\raisebox{0ex}{$\blacktriangle$}~\textcolor{black}{\texttt{+}\hspace{2ex}0.20\small{\%}} & 1.06 & 540.5k & \textbf{31.2} & \color{white}\raisebox{0ex}{$\blacktriangle$}~\textcolor{black}{\texttt{+}\hspace{2ex}2.20\small{\%}} & 1.73 \\
& 6 & 39.0 & 17.8k & 41.4 & \color{white}\raisebox{0ex}{$\blacktriangle$}~\textcolor{black}{\texttt{+}\hspace{2ex}2.40\small{\%}} & 1.60 & 791.4k & \textbf{43.0} & \color{green}\raisebox{0ex}{$\blacktriangle$}~\textcolor{black}{\texttt{+}\hspace{2ex}4.00\small{\%}} & 2.82 \\
& 7 & 47.4 & 17.5k & 66.4 & \color{green}\raisebox{0ex}{$\blacktriangle$}~\textcolor{black}{\texttt{+}\hspace{0.875ex}19.00\small{\%}} & 6.00 & 524.8k & \textbf{74.6} & \color{green}\raisebox{0ex}{$\blacktriangle$}~\textcolor{black}{\texttt{+}\hspace{0.875ex}27.20\small{\%}} & 12.33 \\
& 8 & 34.0 & 17.0k & 35.2 & \color{white}\raisebox{0ex}{$\blacktriangle$}~\textcolor{black}{\texttt{+}\hspace{2ex}1.20\small{\%}} & 1.43 & 850.5k & \textbf{37.4} & \color{green}\raisebox{0ex}{$\blacktriangle$}~\textcolor{black}{\texttt{+}\hspace{2ex}3.40\small{\%}} & 2.06 \\
& 9 & 41.8 & 15.9k & 54.8 & \color{green}\raisebox{0ex}{$\blacktriangle$}~\textcolor{black}{\texttt{+}\hspace{0.875ex}13.00\small{\%}} & 6.00 & 580.9k & \textbf{55.4} & \color{green}\raisebox{0ex}{$\blacktriangle$}~\textcolor{black}{\texttt{+}\hspace{0.875ex}13.60\small{\%}} & 7.18 \\
& 10 & 41.6 & 15.8k & 41.8 & \color{white}\raisebox{0ex}{$\blacktriangle$}~\textcolor{black}{\texttt{+}\hspace{2ex}0.20\small{\%}} & 1.04 & 146.3k & \textbf{42.2} & \color{white}\raisebox{0ex}{$\blacktriangle$}~\textcolor{black}{\texttt{+}\hspace{2ex}0.60\small{\%}} & 1.12 \\
\bottomrule
\multirow{10}{*}{\rotatebox[origin=c]{90}{Spring}} & 1 & 27.4 & 17.9k & 28.0 & \color{white}\raisebox{0ex}{$\blacktriangle$}~\textcolor{black}{\texttt{+}\hspace{2ex}0.60\small{\%}} & 1.16 & 228.6k & \textbf{31.2} & \color{green}\raisebox{0ex}{$\blacktriangle$}~\textcolor{black}{\texttt{+}\hspace{2ex}3.80\small{\%}} & 2.27 \\
& 2 & 24.8 & 16.9k & 25.4 & \color{white}\raisebox{0ex}{$\blacktriangle$}~\textcolor{black}{\texttt{+}\hspace{2ex}0.60\small{\%}} & 1.12 & 188.0k & \textbf{28.2} & \color{green}\raisebox{0ex}{$\blacktriangle$}~\textcolor{black}{\texttt{+}\hspace{2ex}3.40\small{\%}} & 2.13 \\
& 3 & 21.0 & 13.4k & 23.2 & \color{white}\raisebox{0ex}{$\blacktriangle$}~\textcolor{black}{\texttt{+}\hspace{2ex}2.20\small{\%}} & 1.92 & 222.0k & \textbf{25.4} & \color{green}\raisebox{0ex}{$\blacktriangle$}~\textcolor{black}{\texttt{+}\hspace{2ex}4.40\small{\%}} & 3.44 \\
& 4 & 19.0 & 12.7k & \textbf{23.6} & \color{green}\raisebox{0ex}{$\blacktriangle$}~\textcolor{black}{\texttt{+}\hspace{2ex}4.60\small{\%}} & 2.92 & 226.5k & 23.4 & \color{green}\raisebox{0ex}{$\blacktriangle$}~\textcolor{black}{\texttt{+}\hspace{2ex}4.40\small{\%}} & 2.83 \\
& 5 & 28.4 & 12.6k & 28.6 & \color{white}\raisebox{0ex}{$\blacktriangle$}~\textcolor{black}{\texttt{+}\hspace{2ex}0.20\small{\%}} & 1.05 & 223.8k & \textbf{31.8} & \color{green}\raisebox{0ex}{$\blacktriangle$}~\textcolor{black}{\texttt{+}\hspace{2ex}3.40\small{\%}} & 2.00 \\
& 6 & 24.0 & 11.4k & 24.4 & \color{white}\raisebox{0ex}{$\blacktriangle$}~\textcolor{black}{\texttt{+}\hspace{2ex}0.40\small{\%}} & 1.17 & 243.1k & \textbf{27.8} & \color{green}\raisebox{0ex}{$\blacktriangle$}~\textcolor{black}{\texttt{+}\hspace{2ex}3.80\small{\%}} & 2.46 \\
& 7 & 26.0 & 11.2k & 28.0 & \color{white}\raisebox{0ex}{$\blacktriangle$}~\textcolor{black}{\texttt{+}\hspace{2ex}2.00\small{\%}} & 1.83 & 201.5k & \textbf{32.0} & \color{green}\raisebox{0ex}{$\blacktriangle$}~\textcolor{black}{\texttt{+}\hspace{2ex}6.00\small{\%}} & 6.00 \\
& 8 & 21.6 & 10.7k & \textbf{22.6} & \color{white}\raisebox{0ex}{$\blacktriangle$}~\textcolor{black}{\texttt{+}\hspace{2ex}1.00\small{\%}} & 1.42 & 193.0k & 22.0 & \color{white}\raisebox{0ex}{$\blacktriangle$}~\textcolor{black}{\texttt{+}\hspace{2ex}0.40\small{\%}} & 1.11 \\
& 9 & 28.6 & 10.3k & 28.8 & \color{white}\raisebox{0ex}{$\blacktriangle$}~\textcolor{black}{\texttt{+}\hspace{2ex}0.20\small{\%}} & 1.04 & 194.7k & \textbf{29.2} & \color{white}\raisebox{0ex}{$\blacktriangle$}~\textcolor{black}{\texttt{+}\hspace{2ex}0.60\small{\%}} & 1.12 \\
& 10 & 21.6 & 10.1k & 21.8 & \color{white}\raisebox{0ex}{$\blacktriangle$}~\textcolor{black}{\texttt{+}\hspace{2ex}0.20\small{\%}} & 1.05 & 230.8k & \textbf{26.6} & \color{green}\raisebox{0ex}{$\blacktriangle$}~\textcolor{black}{\texttt{+}\hspace{2ex}5.00\small{\%}} & 2.39 \\
\bottomrule
\end{tabular}
}
\end{table*} 
\normalsize
\begin{table*}[htpb] 
    \fontsize{12}{13}\selectfont
    \centering
    \caption{CrystalBLEU (CB) average score between the baseline and the personalized models using T5$_{large}$.}\vspace{-0.2cm}
    \label{tab:results-t5large-cb}
    \vspace*{-0.1cm}
    \resizebox*{!}{0.4\textheight}{
    \rowcolors{2}{gray!20}{white}
    \begin{tabular}{c|c|cccr|cccr}
        \hiderowcolors
        \toprule 
        & \multicolumn{1}{c}{\raisebox{-\heavyrulewidth}{\textbf{~Dev. ID~}}} & \multicolumn{4}{c}{\textbf{Developer}} & \multicolumn{4}{c}{\textbf{Organization}} \\
        \cmidrule{2-10}
        & & \textit{N°} & \textit{CB~\small\%} & \textit{$\Delta$} & \textit{$\vert ES \vert$} & \textit{N°} & \textit{CB~\small\%} & \textit{$\Delta$} & \textit{$\vert ES \vert$} \\
        \midrule
        \multirow{10}{*}{\rotatebox[origin=c]{90}{Apache}} & 1 & 46.6k & 41.66 & \color{green}\raisebox{0ex}{$\blacktriangle$}~\textcolor{black}{\texttt{+}\hspace{0.875ex}18.70\small{\%}} & 0.23 & 888.0k & {46.35} & \color{green}\raisebox{0ex}{$\blacktriangle$}~\textcolor{black}{\texttt{+}\hspace{0.875ex}20.92\small{\%}} & 0.27 \\
        & 2 & 23.2k & 34.73 & \color{green}\raisebox{0ex}{$\blacktriangle$}~\textcolor{black}{\texttt{+}\hspace{2ex}5.82\small{\%}} & 0.09 & 556.0k & {38.62} & \color{green}\raisebox{0ex}{$\blacktriangle$}~\textcolor{black}{\texttt{+}\hspace{0.875ex}11.22\small{\%}} & 0.16 \\
        & 3 & 20.3k & 25.71 & \color{white}\raisebox{0ex}{$\blacktriangledown$}~\textcolor{black}{\texttt{-}\hspace{2ex}0.27\small{\%}} & 0.01 & 830.8k & {29.34} & \color{green}\raisebox{0ex}{$\blacktriangle$}~\textcolor{black}{\texttt{+}\hspace{2ex}4.18\small{\%}} & 0.08 \\
        & 4 & 19.2k & 28.20 & \color{green}\raisebox{0ex}{$\blacktriangle$}~\textcolor{black}{\texttt{+}\hspace{2ex}5.40\small{\%}} & 0.07 & 747.3k & {30.10} & \color{green}\raisebox{0ex}{$\blacktriangle$}~\textcolor{black}{\texttt{+}\hspace{2ex}5.77\small{\%}} & 0.09 \\
        & 5 & 18.6k & 25.35 & \color{white}\raisebox{0ex}{$\blacktriangle$}~\textcolor{black}{\texttt{+}\hspace{2ex}0.82\small{\%}} & 0.02 & 540.5k & {28.61} & \color{green}\raisebox{0ex}{$\blacktriangle$}~\textcolor{black}{\texttt{+}\hspace{2ex}4.49\small{\%}} & 0.07 \\
        & 6 & 17.8k & {29.81} & \color{white}\raisebox{0ex}{$\blacktriangle$}~\textcolor{black}{\texttt{+}\hspace{2ex}4.36\small{\%}} & 0.06 & 791.4k & 29.39 & \color{green}\raisebox{0ex}{$\blacktriangle$}~\textcolor{black}{\texttt{+}\hspace{2ex}6.06\small{\%}} & 0.08 \\
        & 7 & 17.5k & 56.61 & \color{green}\raisebox{0ex}{$\blacktriangle$}~\textcolor{black}{\texttt{+}\hspace{0.875ex}21.78\small{\%}} & 0.32 & 524.8k & {66.42} & \color{green}\raisebox{0ex}{$\blacktriangle$}~\textcolor{black}{\texttt{+}\hspace{0.875ex}33.25\small{\%}} & 0.48 \\
        & 8 & 17.0k & 25.44 & \color{white}\raisebox{0ex}{$\blacktriangle$}~\textcolor{black}{\texttt{+}\hspace{2ex}3.14\small{\%}} & 0.03 & 850.5k & {28.52} & \color{green}\raisebox{0ex}{$\blacktriangle$}~\textcolor{black}{\texttt{+}\hspace{2ex}5.77\small{\%}} & 0.09 \\
        & 9 & 15.9k & 44.47 & \color{green}\raisebox{0ex}{$\blacktriangle$}~\textcolor{black}{\texttt{+}\hspace{0.875ex}19.33\small{\%}} & 0.26 & 580.9k & {45.42} & \color{green}\raisebox{0ex}{$\blacktriangle$}~\textcolor{black}{\texttt{+}\hspace{0.875ex}20.77\small{\%}} & 0.29 \\
        & 10 & 15.8k & 33.39 & \color{white}\raisebox{0ex}{$\blacktriangle$}~\textcolor{black}{\texttt{+}\hspace{2ex}1.08\small{\%}} & 0.03 & 146.3k & {35.36} & \color{white}\raisebox{0ex}{$\blacktriangle$}~\textcolor{black}{\texttt{+}\hspace{2ex}3.48\small{\%}} & 0.06 \\
        \bottomrule
        \multirow{10}{*}{\rotatebox[origin=c]{90}{Spring}} & 1 & 17.9k & 24.18 & \color{white}\raisebox{0ex}{$\blacktriangle$}~\textcolor{black}{\texttt{+}\hspace{2ex}0.64\small{\%}} & 0.02 & 228.6k & {29.45} & \color{green}\raisebox{0ex}{$\blacktriangle$}~\textcolor{black}{\texttt{+}\hspace{2ex}6.71\small{\%}} & 0.11 \\
& 2 & 16.9k & 27.19 & \color{green}\raisebox{0ex}{$\blacktriangle$}~\textcolor{black}{\texttt{+}\hspace{2ex}6.86\small{\%}} & 0.15 & 188.0k & {29.04} & \color{green}\raisebox{0ex}{$\blacktriangle$}~\textcolor{black}{\texttt{+}\hspace{0.875ex}10.54\small{\%}} & 0.20 \\
& 3 & 13.4k & 23.51 & \color{green}\raisebox{0ex}{$\blacktriangle$}~\textcolor{black}{\texttt{+}\hspace{2ex}6.67\small{\%}} & 0.12 & 222.0k & {27.30} & \color{green}\raisebox{0ex}{$\blacktriangle$}~\textcolor{black}{\texttt{+}\hspace{0.875ex}11.08\small{\%}} & 0.19 \\
& 4 & 12.7k & 25.57 & \color{green}\raisebox{0ex}{$\blacktriangle$}~\textcolor{black}{\texttt{+}\hspace{2ex}8.84\small{\%}} & 0.18 & 226.5k & {27.35} & \color{green}\raisebox{0ex}{$\blacktriangle$}~\textcolor{black}{\texttt{+}\hspace{0.875ex}10.62\small{\%}} & 0.21 \\
& 5 & 12.6k & 24.91 & \color{white}\raisebox{0ex}{$\blacktriangle$}~\textcolor{black}{\texttt{+}\hspace{2ex}1.50\small{\%}} & 0.03 & 223.8k & {28.94} & \color{green}\raisebox{0ex}{$\blacktriangle$}~\textcolor{black}{\texttt{+}\hspace{2ex}6.15\small{\%}} & 0.11 \\
& 6 & 11.4k & 19.55 & \color{white}\raisebox{0ex}{$\blacktriangle$}~\textcolor{black}{\texttt{+}\hspace{2ex}1.82\small{\%}} & 0.03 & 243.1k & {28.16} & \color{green}\raisebox{0ex}{$\blacktriangle$}~\textcolor{black}{\texttt{+}\hspace{0.875ex}10.22\small{\%}} & 0.20 \\
& 7 & 11.2k & 22.53 & \color{green}\raisebox{0ex}{$\blacktriangle$}~\textcolor{black}{\texttt{+}\hspace{2ex}3.23\small{\%}} & 0.05 & 201.5k & {26.53} & \color{green}\raisebox{0ex}{$\blacktriangle$}~\textcolor{black}{\texttt{+}\hspace{2ex}8.52\small{\%}} & 0.13 \\
& 8 & 10.7k & 18.93 & \color{white}\raisebox{0ex}{$\blacktriangle$}~\textcolor{black}{\texttt{+}\hspace{2ex}1.05\small{\%}} & 0.03 & 193.0k & {25.12} & \color{green}\raisebox{0ex}{$\blacktriangle$}~\textcolor{black}{\texttt{+}\hspace{2ex}5.84\small{\%}} & 0.15 \\
& 9 & 10.3k & 28.94 & \color{white}\raisebox{0ex}{$\blacktriangle$}~\textcolor{black}{\texttt{+}\hspace{2ex}2.87\small{\%}} & 0.05 & 194.7k & {32.55} & \color{green}\raisebox{0ex}{$\blacktriangle$}~\textcolor{black}{\texttt{+}\hspace{2ex}6.87\small{\%}} & 0.12 \\
& 10 & 10.1k & 21.18 & \color{white}\raisebox{0ex}{$\blacktriangle$}~\textcolor{black}{\texttt{+}\hspace{2ex}1.81\small{\%}} & 0.05 & 230.8k & {28.88} & \color{green}\raisebox{0ex}{$\blacktriangle$}~\textcolor{black}{\texttt{+}\hspace{2ex}9.71\small{\%}} & 0.15 \\
        \bottomrule
        \end{tabular}
        }
        \end{table*} 
        \normalsize

When generating the predictions for the developers' test sets with Code Llama, we observed that the model tended to generate longer outputs, often spanning multiple methods, while our code completion task is capped to at most 50 tokens from the same method (see \secref{sub:dev}). For example, given the masked code shown in Listing~\ref{lst:masked-code}, referring to an \texttt{add} operation, Code Llama may generate a completion like the one shown in Listing~\ref{lst:codellama-output} which, while correct, includes an additional method, \texttt{subtract}. Instead, it should generate only the missing part of the \texttt{add} method, as shown in Listing~\ref{lst:expected-output}. When computing EM predictions, this may lead to an underestimation of the Code Llama capabilities. Thus, in our analysis we also include a version of Code Llama which has been fine-tuned on only 10k instances from random projects (not belonging to Apache or Spring) for one epoch, just to make the model understand the need to generate shorter completions belonging to a single method. We experimentally verified that this small fine-tuning is enough to adapt the model to the task at hand. We refer to this model as B$_{c10k}$.

Table~\ref{tab:results-codellama} shows the results obtained in terms of EM predictions by Code Llama (Baseline B$_{c}$), the version of Code Llama fine-tuned on 10k instances (Baseline B$_{c10k}$), and the \ds and \os models, fine-tuned on top of B$_{c}$. The differences reported for the \ds and \os models (\textit{$\Delta$} \& \textit{OR}) are with respect to B$_{c10k}$ since, as it can be seen, the B$_{c}$ model taken out of the box achieves a low percentage of EM predictions, for the reasons previously explained (\ie it generates additional unrequested methods).

\begin{table*}[htpb] 
    \fontsize{12}{13}\selectfont
    \centering
    \caption{Exact Match (EM) predictions generated by the baseline and by the personalized models using Code Llama 7B. \vspace{-0.3cm}}
    \label{tab:results-codellama}
    \vspace*{-0.1cm}
    \resizebox*{!}{0.4\textheight}{
    \rowcolors{2}{gray!20}{white}
    \begin{tabular}{c|c|cc|cccr|cccr}
        \hiderowcolors
        \toprule 
         & \multicolumn{1}{c}{\raisebox{-\heavyrulewidth}{\textbf{~Dev. ID~}}} & \multicolumn{2}{c}{\raisebox{-\heavyrulewidth}{\textbf{~Baseline~}}} & \multicolumn{4}{c}{\textbf{Developer}} & \multicolumn{4}{c}{\textbf{Organization}} \\
        \cmidrule{2-12}
        & & \textit{EM~\small\% \Large (B$_{c}$)} & \textit{EM~\small\% \Large (B$_{c10k}$)} & \textit{N°} & \textit{EM~\small\%} & \textit{$\Delta$} & \textit{OR} & \textit{N°} & \textit{EM~\small\%} & \textit{$\Delta$} & \textit{OR} \\
        \midrule
        \multirow{10}{*}{\rotatebox[origin=c]{90}{Apache}} & 1 & 8.8 & 62.6 & 46.6k & 72.4 & \color{green}\raisebox{0ex}{$\blacktriangle$}~\textcolor{black}{\texttt{+}\hspace{2ex}9.80\small{\%}} & 9.17 & 888.0k & \textbf{74.0} & \color{green}\raisebox{0ex}{$\blacktriangle$}~\textcolor{black}{\texttt{+}\hspace{0.875ex}11.40\small{\%}} & 15.25 \\
& 2 & 22.6 & 54.4 & 23.2k & \textbf{56.0} & \color{white}\raisebox{0ex}{$\blacktriangle$}~\textcolor{black}{\texttt{+}\hspace{2ex}1.60\small{\%}} & 1.36 & 556.0k & 54.2 & \color{white}\raisebox{0ex}{$\blacktriangledown$}~\textcolor{black}{\texttt{-}\hspace{2ex}0.20\small{\%}} & 0.96 \\
& 3 & 20.0 & 45.6 & 20.3k & 44.8 & \color{white}\raisebox{0ex}{$\blacktriangledown$}~\textcolor{black}{\texttt{-}\hspace{2ex}0.80\small{\%}} & 0.79 & 830.8k & \textbf{45.8} & \color{white}\raisebox{0ex}{$\blacktriangle$}~\textcolor{black}{\texttt{+}\hspace{2ex}0.20\small{\%}} & 1.06 \\
& 4 & 19.6 & 56.8 & 19.2k & 57.8 & \color{white}\raisebox{0ex}{$\blacktriangle$}~\textcolor{black}{\texttt{+}\hspace{2ex}1.00\small{\%}} & 1.25 & 747.3k & \textbf{59.4} & \color{white}\raisebox{0ex}{$\blacktriangle$}~\textcolor{black}{\texttt{+}\hspace{2ex}2.60\small{\%}} & 2.00 \\
& 5 & 22.8 & \textbf{48.6} & 18.6k & \textbf{48.6} & \color{white}\raisebox{0.5ex}{\rule{0.60em}{0.2em}}~\textcolor{black}{\texttt{ }\hspace{2ex}0.00\small{\%}} & 1.00 & 540.5k & 47.4 & \color{white}\raisebox{0ex}{$\blacktriangledown$}~\textcolor{black}{\texttt{-}\hspace{2ex}1.20\small{\%}} & 0.71 \\
& 6 & 16.6 & 49.0 & 17.8k & \textbf{51.8} & \color{green}\raisebox{0ex}{$\blacktriangle$}~\textcolor{black}{\texttt{+}\hspace{2ex}2.80\small{\%}} & 2.75 & 791.4k & 50.2 & \color{white}\raisebox{0ex}{$\blacktriangle$}~\textcolor{black}{\texttt{+}\hspace{2ex}1.20\small{\%}} & 1.38 \\
& 7 & 36.2 & 69.8 & 17.5k & 76.6 & \color{green}\raisebox{0ex}{$\blacktriangle$}~\textcolor{black}{\texttt{+}\hspace{2ex}6.80\small{\%}} & 4.40 & 524.8k & \textbf{80.8} & \color{green}\raisebox{0ex}{$\blacktriangle$}~\textcolor{black}{\texttt{+}\hspace{0.875ex}11.00\small{\%}} & 12.00 \\
& 8 & 16.6 & 46.0 & 17.0k & 47.0 & \color{white}\raisebox{0ex}{$\blacktriangle$}~\textcolor{black}{\texttt{+}\hspace{2ex}1.00\small{\%}} & 1.20 & 850.5k & \textbf{48.6} & \color{white}\raisebox{0ex}{$\blacktriangle$}~\textcolor{black}{\texttt{+}\hspace{2ex}2.60\small{\%}} & 1.93 \\
& 9 & 22.0 & 54.0 & 15.9k & 59.6 & \color{green}\raisebox{0ex}{$\blacktriangle$}~\textcolor{black}{\texttt{+}\hspace{2ex}5.60\small{\%}} & 2.47 & 580.9k & \textbf{60.6} & \color{green}\raisebox{0ex}{$\blacktriangle$}~\textcolor{black}{\texttt{+}\hspace{2ex}6.60\small{\%}} & 5.12 \\
& 10 & 25.2 & 52.6 & 15.8k & 56.2 & \color{green}\raisebox{0ex}{$\blacktriangle$}~\textcolor{black}{\texttt{+}\hspace{2ex}3.60\small{\%}} & 1.67 & 146.3k & \textbf{57.8} & \color{green}\raisebox{0ex}{$\blacktriangle$}~\textcolor{black}{\texttt{+}\hspace{2ex}5.20\small{\%}} & 2.62 \\
        \bottomrule
        \multirow{10}{*}{\rotatebox[origin=c]{90}{Spring}} & 1 & 16.2 & 41.8 & 17.9k & \textbf{44.2} & \color{white}\raisebox{0ex}{$\blacktriangle$}~\textcolor{black}{\texttt{+}\hspace{2ex}2.40\small{\%}} & 1.92 & 228.6k & \textbf{44.2} & \color{green}\raisebox{0ex}{$\blacktriangle$}~\textcolor{black}{\texttt{+}\hspace{2ex}2.40\small{\%}} & 2.71 \\
& 2 & 12.0 & 38.8 & 16.9k & 41.4 & \color{white}\raisebox{0ex}{$\blacktriangle$}~\textcolor{black}{\texttt{+}\hspace{2ex}2.60\small{\%}} & 1.93 & 188.0k & \textbf{42.2} & \color{green}\raisebox{0ex}{$\blacktriangle$}~\textcolor{black}{\texttt{+}\hspace{2ex}3.40\small{\%}} & 3.12 \\
& 3 & 13.0 & 35.6 & 13.4k & 37.4 & \color{white}\raisebox{0ex}{$\blacktriangle$}~\textcolor{black}{\texttt{+}\hspace{2ex}1.80\small{\%}} & 1.50 & 222.0k & \textbf{37.8} & \color{white}\raisebox{0ex}{$\blacktriangle$}~\textcolor{black}{\texttt{+}\hspace{2ex}2.20\small{\%}} & 1.85 \\
& 4 & 10.0 & 32.8 & 12.7k & \textbf{35.6} & \color{white}\raisebox{0ex}{$\blacktriangle$}~\textcolor{black}{\texttt{+}\hspace{2ex}2.80\small{\%}} & 1.82 & 226.5k & 35.2 & \color{white}\raisebox{0ex}{$\blacktriangle$}~\textcolor{black}{\texttt{+}\hspace{2ex}2.40\small{\%}} & 1.75 \\
& 5 & 13.4 & 43.4 & 12.6k & 45.8 & \color{white}\raisebox{0ex}{$\blacktriangle$}~\textcolor{black}{\texttt{+}\hspace{2ex}2.40\small{\%}} & 2.09 & 223.8k & \textbf{46.8} & \color{green}\raisebox{0ex}{$\blacktriangle$}~\textcolor{black}{\texttt{+}\hspace{2ex}3.40\small{\%}} & 2.70 \\
& 6 & 8.6 & 36.8 & 11.4k & \textbf{39.4} & \color{white}\raisebox{0ex}{$\blacktriangle$}~\textcolor{black}{\texttt{+}\hspace{2ex}2.60\small{\%}} & 2.00 & 243.1k & 39.2 & \color{white}\raisebox{0ex}{$\blacktriangle$}~\textcolor{black}{\texttt{+}\hspace{2ex}2.40\small{\%}} & 1.71 \\
& 7 & 15.6 & 38.0 & 11.2k & 42.2 & \color{green}\raisebox{0ex}{$\blacktriangle$}~\textcolor{black}{\texttt{+}\hspace{2ex}4.20\small{\%}} & 2.40 & 201.5k & \textbf{44.0} & \color{green}\raisebox{0ex}{$\blacktriangle$}~\textcolor{black}{\texttt{+}\hspace{2ex}6.00\small{\%}} & 3.50 \\
& 8 & 12.8 & 39.8 & 10.7k & 40.2 & \color{white}\raisebox{0ex}{$\blacktriangle$}~\textcolor{black}{\texttt{+}\hspace{2ex}0.40\small{\%}} & 1.08 & 193.0k & \textbf{43.2} & \color{green}\raisebox{0ex}{$\blacktriangle$}~\textcolor{black}{\texttt{+}\hspace{2ex}3.40\small{\%}} & 2.06 \\
& 9 & 21.6 & 47.2 & 10.3k & 49.6 & \color{white}\raisebox{0ex}{$\blacktriangle$}~\textcolor{black}{\texttt{+}\hspace{2ex}2.40\small{\%}} & 1.60 & 194.7k & \textbf{50.8} & \color{green}\raisebox{0ex}{$\blacktriangle$}~\textcolor{black}{\texttt{+}\hspace{2ex}3.60\small{\%}} & 1.78 \\
& 10 & 13.0 & 35.6 & 10.1k & 38.6 & \color{green}\raisebox{0ex}{$\blacktriangle$}~\textcolor{black}{\texttt{+}\hspace{2ex}3.00\small{\%}} & 1.88 & 230.8k & \textbf{43.4} & \color{green}\raisebox{0ex}{$\blacktriangle$}~\textcolor{black}{\texttt{+}\hspace{2ex}7.80\small{\%}} & 4.00 \\
\bottomrule
\end{tabular}
}
\end{table*} 
\normalsize
\begin{table*}[htpb] 
    \fontsize{12}{13}\selectfont
    \centering
    \caption{CrystalBLEU (CB) average score between the baseline and the personalized Code Llama 7B models.\vspace{-0.3cm}}
    \label{tab:results-codellama-cb}
    \vspace*{-0.1cm}
    \resizebox*{!}{0.4\textheight}{
    \rowcolors{2}{gray!20}{white}
    \begin{tabular}{c|c|cccr|cccr}
        \hiderowcolors
        \toprule 
        & \multicolumn{1}{c}{\raisebox{-\heavyrulewidth}{\textbf{~Dev. ID~}}} & \multicolumn{4}{c}{\textbf{Developer}} & \multicolumn{4}{c}{\textbf{Organization}} \\
        \cmidrule{2-10}
        & & \textit{N°} & \textit{CB~\small\%} & \textit{$\Delta$} & \textit{$\vert ES \vert$} & \textit{N°} & \textit{CB~\small\%} & \textit{$\Delta$} & \textit{$\vert ES \vert$} \\
        \midrule
        \multirow{10}{*}{\rotatebox[origin=c]{90}{Apache}} & 1 & 46.6k & 44.58 & \color{green}\raisebox{0ex}{$\blacktriangle$}~\textcolor{black}{\texttt{+}\hspace{0.875ex}18.39\small{\%}} & 0.22 & 888.0k & {47.72} & \color{green}\raisebox{0ex}{$\blacktriangle$}~\textcolor{black}{\texttt{+}\hspace{0.875ex}22.31\small{\%}} & 0.27 \\
& 2 & 23.2k & 38.50 & \color{white}\raisebox{0ex}{$\blacktriangle$}~\textcolor{black}{\texttt{+}\hspace{2ex}2.52\small{\%}} & 0.03 & 556.0k & {38.73} & \color{white}\raisebox{0ex}{$\blacktriangle$}~\textcolor{black}{\texttt{+}\hspace{2ex}2.24\small{\%}} & 0.04 \\
& 3 & 20.3k & {33.19} & \color{white}\raisebox{0ex}{$\blacktriangle$}~\textcolor{black}{\texttt{+}\hspace{2ex}0.56\small{\%}} & 0.01 & 830.8k & 31.95 & \color{white}\raisebox{0ex}{$\blacktriangle$}~\textcolor{black}{\texttt{+}\hspace{2ex}0.02\small{\%}} & 0.01 \\
& 4 & 19.2k & 37.31 & \color{white}\raisebox{0ex}{$\blacktriangle$}~\textcolor{black}{\texttt{+}\hspace{2ex}2.60\small{\%}} & 0.04 & 747.3k & {37.41} & \color{white}\raisebox{0ex}{$\blacktriangle$}~\textcolor{black}{\texttt{+}\hspace{2ex}4.69\small{\%}} & 0.06 \\
& 5 & 18.6k & {34.16} & \color{white}\raisebox{0ex}{$\blacktriangle$}~\textcolor{black}{\texttt{+}\hspace{2ex}0.72\small{\%}} & 0.02 & 540.5k & 33.53 & \color{white}\raisebox{0ex}{$\blacktriangledown$}~\textcolor{black}{\texttt{-}\hspace{2ex}0.15\small{\%}} & 0.00 \\
& 6 & 17.8k & {34.68} & \color{green}\raisebox{0ex}{$\blacktriangle$}~\textcolor{black}{\texttt{+}\hspace{2ex}4.51\small{\%}} & 0.06 & 791.4k & 32.60 & \color{white}\raisebox{0ex}{$\blacktriangle$}~\textcolor{black}{\texttt{+}\hspace{2ex}0.37\small{\%}} & 0.01 \\
& 7 & 17.5k & 48.18 & \color{green}\raisebox{0ex}{$\blacktriangle$}~\textcolor{black}{\texttt{+}\hspace{0.875ex}12.39\small{\%}} & 0.17 & 524.8k & {56.10} & \color{green}\raisebox{0ex}{$\blacktriangle$}~\textcolor{black}{\texttt{+}\hspace{0.875ex}22.37\small{\%}} & 0.32 \\
& 8 & 17.0k & {36.40} & \color{white}\raisebox{0ex}{$\blacktriangle$}~\textcolor{black}{\texttt{+}\hspace{2ex}1.51\small{\%}} & 0.01 & 850.5k & 35.71 & \color{green}\raisebox{0ex}{$\blacktriangle$}~\textcolor{black}{\texttt{+}\hspace{2ex}3.34\small{\%}} & 0.04 \\
& 9 & 15.9k & {41.45} & \color{green}\raisebox{0ex}{$\blacktriangle$}~\textcolor{black}{\texttt{+}\hspace{2ex}8.44\small{\%}} & 0.12 & 580.9k & 40.84 & \color{green}\raisebox{0ex}{$\blacktriangle$}~\textcolor{black}{\texttt{+}\hspace{0.875ex}10.92\small{\%}} & 0.16 \\
& 10 & 15.8k & {46.83} & \color{green}\raisebox{0ex}{$\blacktriangle$}~\textcolor{black}{\texttt{+}\hspace{2ex}7.35\small{\%}} & 0.11 & 146.3k & 45.77 & \color{green}\raisebox{0ex}{$\blacktriangle$}~\textcolor{black}{\texttt{+}\hspace{2ex}8.92\small{\%}} & 0.15 \\
        \bottomrule
        \multirow{10}{*}{\rotatebox[origin=c]{90}{Spring}} & 1 & 17.9k & {34.48} & \color{green}\raisebox{0ex}{$\blacktriangle$}~\textcolor{black}{\texttt{+}\hspace{2ex}3.96\small{\%}} & 0.07 & 228.6k & 33.08 & \color{green}\raisebox{0ex}{$\blacktriangle$}~\textcolor{black}{\texttt{+}\hspace{2ex}3.96\small{\%}} & 0.07 \\
& 2 & 16.9k & 34.79 & \color{green}\raisebox{0ex}{$\blacktriangle$}~\textcolor{black}{\texttt{+}\hspace{2ex}4.92\small{\%}} & 0.08 & 188.0k & {36.26} & \color{green}\raisebox{0ex}{$\blacktriangle$}~\textcolor{black}{\texttt{+}\hspace{2ex}7.73\small{\%}} & 0.12 \\
& 3 & 13.4k & {33.94} & \color{white}\raisebox{0ex}{$\blacktriangle$}~\textcolor{black}{\texttt{+}\hspace{2ex}2.54\small{\%}} & 0.03 & 222.0k & 33.60 & \color{green}\raisebox{0ex}{$\blacktriangle$}~\textcolor{black}{\texttt{+}\hspace{2ex}3.23\small{\%}} & 0.05 \\
& 4 & 12.7k & {33.59} & \color{green}\raisebox{0ex}{$\blacktriangle$}~\textcolor{black}{\texttt{+}\hspace{2ex}5.82\small{\%}} & 0.09 & 226.5k & 33.50 & \color{green}\raisebox{0ex}{$\blacktriangle$}~\textcolor{black}{\texttt{+}\hspace{2ex}5.93\small{\%}} & 0.10 \\
& 5 & 12.6k & 34.03 & \color{green}\raisebox{0ex}{$\blacktriangle$}~\textcolor{black}{\texttt{+}\hspace{2ex}4.58\small{\%}} & 0.07 & 223.8k & {36.39} & \color{green}\raisebox{0ex}{$\blacktriangle$}~\textcolor{black}{\texttt{+}\hspace{2ex}7.18\small{\%}} & 0.11 \\
& 6 & 11.4k & 32.64 & \color{green}\raisebox{0ex}{$\blacktriangle$}~\textcolor{black}{\texttt{+}\hspace{2ex}6.90\small{\%}} & 0.12 & 243.1k & {34.61} & \color{green}\raisebox{0ex}{$\blacktriangle$}~\textcolor{black}{\texttt{+}\hspace{2ex}7.98\small{\%}} & 0.14 \\
& 7 & 11.2k & {36.84} & \color{green}\raisebox{0ex}{$\blacktriangle$}~\textcolor{black}{\texttt{+}\hspace{2ex}6.64\small{\%}} & 0.11 & 201.5k & 35.84 & \color{green}\raisebox{0ex}{$\blacktriangle$}~\textcolor{black}{\texttt{+}\hspace{2ex}6.29\small{\%}} & 0.09 \\
& 8 & 10.7k & 33.33 & \color{white}\raisebox{0ex}{$\blacktriangle$}~\textcolor{black}{\texttt{+}\hspace{2ex}3.09\small{\%}} & 0.07 & 193.0k & {35.40} & \color{green}\raisebox{0ex}{$\blacktriangle$}~\textcolor{black}{\texttt{+}\hspace{2ex}6.92\small{\%}} & 0.11 \\
& 9 & 10.3k & 39.29 & \color{green}\raisebox{0ex}{$\blacktriangle$}~\textcolor{black}{\texttt{+}\hspace{2ex}7.18\small{\%}} & 0.11 & 194.7k & {42.37} & \color{green}\raisebox{0ex}{$\blacktriangle$}~\textcolor{black}{\texttt{+}\hspace{2ex}9.55\small{\%}} & 0.15 \\
& 10 & 10.1k & 32.50 & \color{green}\raisebox{0ex}{$\blacktriangle$}~\textcolor{black}{\texttt{+}\hspace{2ex}6.74\small{\%}} & 0.11 & 230.8k & {36.69} & \color{green}\raisebox{0ex}{$\blacktriangle$}~\textcolor{black}{\texttt{+}\hspace{0.875ex}11.82\small{\%}} & 0.17 \\
\bottomrule
\end{tabular}
}
\end{table*} 
\normalsize

As illustrated, personalization is effective also for Code Llama. When it comes to the \ds models, eight out of 10 Apache developers and all 10 Spring developers observed a boost in performance. For Apache, this is statistically significant for five developers, with an average increase in EM predictions of 5.72\% and an average OR of 4.09 (min=2.47, max=9.17). As per Spring, two developers have a statistically significant improvement of EM, with a +4.2\% and a +3.0\% of EM predictions (ORs 2.40 and 1.88, respectively). As already observed for the T5 models, it is the \os fine-tuning that brings most of the performance improvement (see \tabref{tab:results-codellama}): Overall, for 11 out of 20 developers (four for Apache and seven for Spring) Code Llama got a statistically significant boost in EM predictions, with an average increase of 5.84\% and an average OR of 4.99 (min=2.06, max=15.25).

The findings are consistent also when looking at the CrystalBLEU scores (\tabref{tab:results-codellama-cb}). Still focusing on the most successful personalization (\ie \os), a statistically significant increase in CrystalBLEU is observed on the test sets of 5/10 Apache and 10/10 Spring developers. For four of these developers, a two-digit increase is observed, indicating a strong impact of the personalized fine-tuning.

\begin{tcolorbox}[title=\faLightbulbO~~Summary of Findings]
    Personalization is effective across different model sizes (60M, 750M and 7B parameters), and model architectures (T5 and Llama-based). Even Code Llama that likely saw code from the two subject organizations at training time benefitted of further specialized fine-tuning.
\end{tcolorbox}

\subsection{Goal 4: Investigating the Cost-Performance Trade-Off}
\label{sec:cost-analysis}

As explained in \secref{sec:cost-effectiveness}, we also run an additional analysis aimed at understanding the cost-effectiveness of the personalized fine-tuning. Indeed, the personalized fine-tuning implies a training cost that the company would not have by just downloading a larger code-completion model already trained for such a task, and maybe exhibiting even better performance than the smaller, personalized model. For the reasons detailed in \secref{sec:cost-effectiveness}, we perform this analysis between the generic T5$_{large}$ model (\ie B$_{l}$), as represenattive of a general-purpose model that a company could download and use out of the box, and the personalized \os T5$_{small}$ models.

\begin{table*}[t] 
    \fontsize{12}{13}\selectfont
    \centering
    \caption{Exact Match (EM) predictions of the base T5$_{large}$ \emph{vs} T5$_{small}$ \os models.\vspace{-0.3cm}}
    \label{tab:results-t5large-vs-small}
    \resizebox*{!}{0.4\textheight}{
    \rowcolors{2}{gray!20}{white}
    \begin{tabular}{c|c|c|cccr}
        \hiderowcolors
        \toprule 
        & \multicolumn{1}{c}{\raisebox{-\heavyrulewidth}{\textbf{~Dev. ID~}}} & \multicolumn{1}{c}{\raisebox{-\heavyrulewidth}{\textbf{\ T5$_{large}$\ }}} & \multicolumn{4}{c}{\textbf{T5$_{small}$ organization}} \\
        \cmidrule{2-7}
        & & \textit{EM~\small\%} & \textit{N°} & \textit{EM~\small\%} & \textit{$\Delta$} & \textit{OR} \\
        \midrule
\multirow{10}{*}{\rotatebox[origin=c]{90}{Apache}} & 1 & 52.8 & 888.0k & 61.8 & \color{green}\raisebox{0ex}{$\blacktriangle$}~\textcolor{black}{\texttt{+}\hspace{2ex}9.00\small{\%}} & 3.81 \\
& 2 & 37.4 & 556.0k & 37.0 & \color{white!20}\raisebox{0ex}{$\blacktriangledown$}~\textcolor{black}{\texttt{-}\hspace{2ex}0.40\small{\%}} & 0.94 \\
& 3 & 30.6 & 830.8k & 28.2 & \color{white}\raisebox{0ex}{$\blacktriangledown$}~\textcolor{black}{\texttt{-}\hspace{2ex}2.40\small{\%}} & 0.56 \\
& 4 & 42.6 & 747.3k & 44.2 & \color{white}\raisebox{0ex}{$\blacktriangle$}~\textcolor{black}{\texttt{+}\hspace{2ex}1.60\small{\%}} & 1.38 \\
& 5 & 29.0 & 540.5k & 24.0 & \color{red}\raisebox{0ex}{$\blacktriangledown$}~\textcolor{black}{\texttt{-}\hspace{2ex}5.00\small{\%}} & 0.42 \\
& 6 & 39.0 & 791.4k & 37.2 & \color{white}\raisebox{0ex}{$\blacktriangledown$}~\textcolor{black}{\texttt{-}\hspace{2ex}1.80\small{\%}} & 0.67 \\
& 7 & 47.4 & 524.8k & 70.0 & \color{green}\raisebox{0ex}{$\blacktriangle$}~\textcolor{black}{\texttt{+}\hspace{0.875ex}22.60\small{\%}} & 6.95 \\
& 8 & 34.0 & 850.5k & 32.2 & \color{white}\raisebox{0ex}{$\blacktriangledown$}~\textcolor{black}{\texttt{-}\hspace{2ex}1.80\small{\%}} & 0.68 \\
& 9 & 41.8 & 580.9k & 48.4 & \color{green}\raisebox{0ex}{$\blacktriangle$}~\textcolor{black}{\texttt{+}\hspace{2ex}6.60\small{\%}} & 2.14 \\
& 10 & 41.6 & 146.3k & 33.6 & \color{red}\raisebox{0ex}{$\blacktriangledown$}~\textcolor{black}{\texttt{-}\hspace{2ex}8.00\small{\%}} & 0.30 \\
\bottomrule
\multirow{10}{*}{\rotatebox[origin=c]{90}{Spring}} & 1 & 27.4 & 228.6k & 27.4 & \color{white}\raisebox{0.5ex}{\rule{0.60em}{0.2em}}~\textcolor{black}{\texttt{ }\hspace{2ex}0.00\small{\%}} & 1.00 \\
& 2 & 24.8 & 188.0k & 26.0 & \color{white}\raisebox{0ex}{$\blacktriangle$}~\textcolor{black}{\texttt{+}\hspace{2ex}1.20\small{\%}} & 1.35 \\
& 3 & 21.0 & 222.0k & 21.0 & \color{white}\raisebox{0.5ex}{\rule{0.60em}{0.2em}}~\textcolor{black}{\texttt{ }\hspace{2ex}0.00\small{\%}} & 1.00 \\
& 4 & 19.0 & 226.5k & 19.4 & \color{white}\raisebox{0ex}{$\blacktriangle$}~\textcolor{black}{\texttt{+}\hspace{2ex}0.40\small{\%}} & 1.12 \\
& 5 & 28.4 & 223.8k & 28.2 & \color{white}\raisebox{0ex}{$\blacktriangledown$}~\textcolor{black}{\texttt{-}\hspace{2ex}0.20\small{\%}} & 0.96 \\
& 6 & 24.0 & 243.1k & 25.6 & \color{white}\raisebox{0ex}{$\blacktriangle$}~\textcolor{black}{\texttt{+}\hspace{2ex}1.60\small{\%}} & 1.38 \\
& 7 & 26.0 & 201.5k & 27.8 & \color{white}\raisebox{0ex}{$\blacktriangle$}~\textcolor{black}{\texttt{+}\hspace{2ex}1.80\small{\%}} & 1.56 \\
& 8 & 21.6 & 193.0k & 22.2 & \color{white}\raisebox{0ex}{$\blacktriangle$}~\textcolor{black}{\texttt{+}\hspace{2ex}0.60\small{\%}} & 1.13 \\
& 9 & 28.6 & 194.7k & 26.4 & \color{white}\raisebox{0ex}{$\blacktriangledown$}~\textcolor{black}{\texttt{-}\hspace{2ex}2.20\small{\%}} & 0.70 \\
& 10 & 21.6 & 230.8k & 23.8 & \color{white}\raisebox{0ex}{$\blacktriangle$}~\textcolor{black}{\texttt{+}\hspace{2ex}2.20\small{\%}} & 1.44 \\
\bottomrule
\end{tabular}
}
\end{table*} 
\normalsize

We start by comparing the performance of both models for the top-10 developers of both organizations. \tabref{tab:results-t5large-vs-small} shows the results. As it can be seen, the performance of the two models is comparable on the top-10 developers of both organizations. Indeed, when it comes to Apache, there is a statistically significant difference in EM predictions for five out of 10 developers, three times in favor of T5$_{small}$ and two in favor of T5$_{large}$. As per Spring, the two models are basically equivalent (\ie no statistically significant difference) for all developers. Overall, we can conclude that a model fine-tuned on \os data can be as effective as a generic model that is over 10 times larger (60M \emph{vs} 750M parameters).

In addition to the performance dimension of the comparison, we also aim to understand whether the cost of training a smaller personalized model is justified. Indeed, while the ``generic'' T5$_{large}$ (as well as any other already trained code model) could be just downloaded and used out of the box without any cost, the personalized T5$_{small}$ requires a fine-tuning which has a cost. However, we expect the T5$_{large}$ to have a higher ``inference cost'' (\ie the cost of generating one prediction is higher). \secref{sec:cost-effectiveness} details how we computed the training and inference costs for the models. Remember that for the personalized T5$_{small}$, we consider the training cost of both the cheapest (146.3k training instances) and the most expensive (888k training instances) \os T5$_{small}$ from our study as a sort of lower- and upper-bounds.

\begin{figure}
    \centering
    \includegraphics[width=0.6\textwidth]{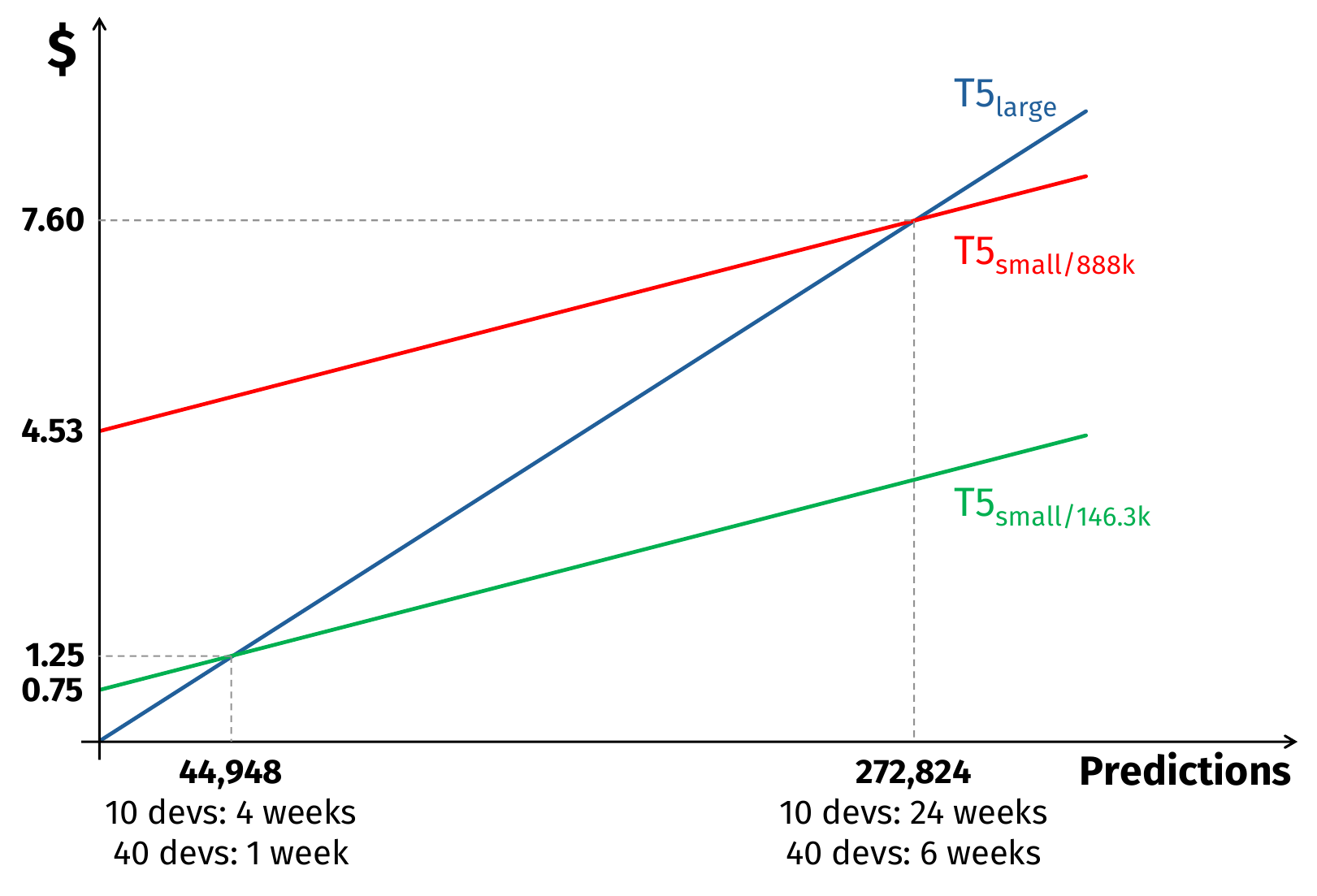}
    \caption{Cost-effectiveness analysis: Generic T5$_{large}$ \emph{vs} \os T5$_{small}$.}
    \label{fig:cost-effectiveness}
\end{figure}

\figref{fig:cost-effectiveness} shows on the $y$ axis the GPU renting cost for all three models (cheapest and most expensive \emph{organization-specific} T5$_{small}$ and the generic T5$_{large}$) given a different number of performed inferences ($x$ axis). When no inferences are performed, T5$_{large}$ costs, in our simulation, 0\$ since we are assuming this is a generic model that the company downloaded and used out of the box. The T5$_{small}$ models, instead, bring with them the training cost (from a minimum of 0.75\$ to a maximum of 4.53\$). As we can see, there is a breakeven point after at least 44,948 and at most 272,824 inferences (best- and worst-case scenario). This leads to the following question: \emph{In how much time do the developers of an organization  reach this number of code completion inferences?} From an internal study at Microsoft \cite{copilot-weekly}, we know that the average number of weekly recommendations that Copilot triggers to a single developer is 1,150. This means that if we consider a software company employing 10 developers, the breakeven point will be reached after four (best-case) to 24 (worst-case) weeks. With 40 developers, this goes down to 1-6 weeks.

\begin{tcolorbox}[title=\faLightbulbO~~Summary of Findings]
    Thanks to a personalized fine-tuning, a company could deploy a $\sim$10$\times$ smaller model being equivalent in terms of code completion performance to the larger model. In terms of costs, the breakeven point is reached in few weeks (depending on the number of developers employed and the size of the used fine-tuning dataset).
\end{tcolorbox}

	\section{Summary of Main Findings}
\label{sec:findings}

We summarize in the following the main findings output of our study, discussing their implications for researchers and software companies.

\begin{enumerate}

\item \emph{The boost in performance obtained by developer-specific models is often not sufficient to justify the additional effort of collecting the needed training data and run the fine-tuning (Goal 1).} The main issue with a developer-specific fine-tuning is the lack of training data for most of developers. This has been observed even on a large organization such as Apache, for which hundreds of long-lived repositories were mined. Thus, we expect the lack of developer-specific data to be a show-stopper for most companies. However, our analysis on the impact of the training data size from \secref{sub:trainingSize} also showed that, given a fixed amount of training data---generic, organization-specific, or developer-specific---there is a clear ranking in their effectiveness: the more specific the training data, the better the provided boost in performance, with the developer-specific ones being the most ``precious''. Basically, assuming the possibility to increase the size of the developer-specific training sets, such a personalized fine-tuning could be even more successful than the organization-specific one. Researchers could look into strategies such as data augmentation to address this limitation. On top of this, in our study we did not look into the performance provided by a combination of an organization-specific fine-tuning \emph{followed} by a developer-specific fine-tuning. Also this strategy could bring benefits that we did not investigate.

\item \emph{An organization-specific fine-tuning should be the obvious choice for most companies interested in deploying an in-house code completion model (Goal 1).} As explained, our analyses showed that this is due to the much higher number of training instances that can be collected at ``organization-level''. While the training cost is much higher than the one of the developer-specific dataset, a single fine-tuning would be required in a real scenario, while the developer-specific fine-tuning requires the training of a different model for each developer. Also, it is more convenient to deploy and maintain a single model.

\item \emph{The increase in performance observed with both specializations is not simply due to a higher number of training instances as compared to the baselines (Goal 2).} Indeed, by further fine-tuning the baselines on generic data (and not on organization/developer-specific data) we did not observe an increase in performance comparable to the one provided by the two specializations---see \secref{sub:trainingSize}. This result stresses the key role played by the specificity of training data, and calls for additional investigation pertaining other code-related tasks (\eg automated bug-fixing, code review).

\item \emph{The increase in performance ensured by personalization can be generalized to models having a different architecture and size (Goal 3)}. Indeed, all models we experimented with (T5 and Code Llama) benefitted from major performance improvements, independently from their size. This is a major finding of our study, since we can conjecture that even the code completion performance of very large models could be further boosted via personalization. Also, the inclusion of Code Llama in our experiments demonstrated that personalization can also work in scenarios in which code from the target organization was likely already seen at pre-training time.

\item \emph{Fine-tuning organization-specific DL models can be cost-effective (Goal 4).} Indeed, we show that thanks to a personalized fine-tuning, DL models can achieve code completion performance on par with those of models being $\sim$10$\times$ larger (\eg an organization-specific T5$_{small}$ achieves the same performance of a ``generic'' T5$_{large}$ model). The lower inference costs of smaller models will allow the companies to save money in the long run, since only a few weeks are needed to reach a breakeven point and amortize the training cost, depending on the number of developers employed in the company---see detailed analysis in \secref{sec:cost-analysis}.

\end{enumerate}
\section{Threats to Validity} \label{sec:threats}

\textbf{Construct validity.} One threat is related to how we assess the code completion performance. While an EM prediction is likely to be useful for developers, it is difficult to speculate about non-EM predictions. The latter may be valuable while still being different from the expected target (\eg the recommended code is different but semantically equivalent). We partially address this threat by also employing the CrystalBLEU as an evaluation metric, to at least provide a measure of how far the prediction is from the target. 

\textbf{Internal validity.} As a design choice, we adopted the original architecture and hyperparameters for all models subject of our study. This was due to the high number of models we had to train for our study (\trainedModels), which would have further increased with hyperparameters tuning. However, since we compared the baseline and the personalized models when using the same exact configuration, we expect no major impact of this choice on our main findings.

Another concern may be related to the effectiveness of the trainings we performed, especially when looking at the models we trained from scratch (\ie the T5 models). While it is difficult to make a fair comparison with other T5 models from the literature which have been trained on different datasets and evaluated on different test sets, one possible point of reference about the ``expected'' performance of T5 for code completion comes from the work by Ciniselli \etal \cite{ciniselli:tse2021}, in which the authors experiment the T5$_{small}$ in three different completion scenarios, namely \emph{token}-masking (\ie completing the last $n$ tokens of a statement, with $n$ capped to 10), \emph{construct}-masking (\ie predicting specific code constructs, such as the conditions of \texttt{if} statements), and \emph{block}-masking (\ie predicting up to two complete code statements). Our test sets feature completions masking up to 50 tokens, possibly spanning across multiple statements, thus being ``similar'' to their \emph{block}-masking scenario. When using pre-training and single-task fine-tuning (\ie the same training procedure we adopt), Ciniselli \etal achieved 27.2\% of EM predictions in the \emph{block}-masking scenario on Java code. On average, across the 136 developers considered in our study, our baseline T5$_{small}$ (no personalized fine-tuning) achieved 26.4\% of EM predictions, thus being aligned with previous findings \cite{ciniselli:tse2021}. This provides some confidence about the correctness of the training procedure.

\textbf{External validity.} While we considered a fairly high number of developers for our study (136), we focused on two organizations and one programming language (Java). Also, we used T5 and Code Llama as representative DL models for code completion. Our findings may not generalize to other settings.
\section{Related Work} \label{sec:related}

Several DL-based techniques have been proposed to improve the automation provided by code completion tools (see \eg \cite{liu:ase2020,svyatkovskiy:msr2020,Ciniselli:msr2021,ciniselli:tse2021,izadi:icse2022,nguyen:tse2022}). Differently from our work, these studies focus on training the model with ``general'' coding knowledge provided via large-scale datasets. Other recent studies propose alternative approaches by leveraging in-context learning abilities of large language models \cite{bairi2023codeplan, zhang2023repocoder, ahmed:icse2024}.
In this section, we mainly discuss studies centered on the customization of recommender systems for code generation tasks and empirical investigations looking at code recommender tools from other perspectives \cite{muaruasoiu:ppig2015,jin:msr2018,hellendoorn:icse2019,liu:tse2020,ciniselli:tse2021,ciniselli:msr2022,xu:tosem2022,ziegler:maps2022,imai:icse2022,Mastropaolo:icse2023,izadi:icse2024}, focusing on reported findings which may relate to the motivations and outcome of our study.

\subsection{Personalized Source Code Recommendations}
To the best of our knowledge, only few works in the literature targeted the personalization of code recommendations. 

Saraiva \etal \cite{saraiva2015products} conducted a preliminary study on the performance of $n$-gram language models on the Microsoft Office suite using three levels of personalization: application-specific, developer-specific, and time-specific. Their findings show that $n$-gram models trained on a single project always perform better than a general model working on the entire Office source code or a specific developer corpus. Furthermore, they found that models built on specific time-based datasets do not lead to particular performance improvements. In this work, we applied a similar idea to DL-based code completion, being the state of the practice nowadays.

Allamanis \etal \cite{allamanis:fse2014} presented \textsc{Naturalize}, a  framework recommending ``\emph{natural}'' identifier names and formatting choices which have been learned from a given codebase, thus improving the stylistic consistency of the project. Our work embraces the basic idea proposed in this work, looking however at the impact of personalization on the code completion capabilities of DL models.

Ahmed and Devanbu \etal \cite{ahmed2022few} investigated the usage of few-shot learning for customizing code summarization tasks. The authors collected code and summary pairs from eight repositories \cite{lu:nips2021}. Then, they evaluated the few-shot capabilities of the model when preceding each query with 10 examples of the same project. They found that using project-specific examples instead of cross-project instances can improve the inference accuracy of the model.

The most related work to our study is the one recently presented by Zlotchevski \etal \cite{Zlotchevski:fse2022}, who studied whether the automated generation of tests can be improved by further fine-tuning a DL model on a specific software repository. Given a DL model already fine-tuned to generate unit tests, the authors further trained it on code coming from a specific code repository, reporting an improved accuracy of the suggestions. In this work, we focus on the more generic task of DL-based code completion, which is used by millions of developers and thousands of companies \cite{microsoft} via tools such as Copilot. Also, we look at different levels of personalization, related to a whole organization and a single developer, which have not been previously explored in the literature.

\subsection{Empirical Studies on Code Recommender Systems}
M{\u{a}}r{\u{a}}șoiu \etal \cite{muaruasoiu:ppig2015} studied how developers use code completion tools, showing that they make extensive use of these tools but often discard the provided suggestions, especially when they feature APIs they are not familiar to. This helps motivating the need for personalized code completion.

Hellendoorn \etal \cite{hellendoorn:icse2019} reported the inadequacy of artificial benchmarks in evaluating code completion tools, since these do not reflect the complexity of real-world completions. Similar findings were reported by Liu \etal \cite{liu:tse2020} for the task of requirements implementation (\ie generating code starting from a natural language description). These findings are among the reasons why our test sets feature code completions derived by real code changes implemented by software developers.

Ciniselli \etal \cite{ciniselli:tse2021} showed that T5$_{small}$ can accurately recommend code completions spanning across a single statement ($\sim$69\% of accuracy) and still achieve good results when dealing with the completion of entire statements ($\sim$29\%). This  motivated our selection of the subject DL model. 

Other studies focused on the effectiveness of the support provided by code completion tools to developers \cite{xu:tosem2022,imai:icse2022,Mastropaolo:icse2023}, documenting limitations of these tools, including: the lack of improvement in developers' productivity \cite{xu:tosem2022}, the presence of low quality recommendations \cite{imai:icse2022}, and issues related to their robustness \cite{Mastropaolo:icse2023}. These works point to the need for additional research aimed at improving code recommenders.
\section{Conclusion and Future Work} \label{sec:conclusion}

In this work, we investigate how personalizing DL-based code completion tools can help in boosting their performance. We show that, by fine-tuning a generic code completion model on \emph{personalized} instances (\ie completions from the same developer/organization), it can achieve significantly better performance, also when compared to a model fine-tuned on the same number of non-personalized instances. Our results hold along four dimensions: (i)~different levels of personalization (\os and \ds data), with \os personalization being, however, more effective thanks to the availability of more training data; (ii)~different developers (136) from two different organizations (Apache and Spring); (iii)~different sizes of the training  dataset used to personalize the model (between 1k and 908.1k instances); and (iv)~different model sizes and architecture (60M and 750M T5 models and 7B Code Llama).

Our findings show that most developers within an organization can benefit from personalized recommendations, even when no \ds data is available (\eg using the \os model for developers new to the organization).
Furthermore, companies providing industry-ready code completion tools (\eg GitHub Copilot~\cite{copilot}) may find new business opportunities by offering personalized models to their customers, while smaller-scale models deployed in-house, when personalized, may be competitive and closer to what offered by larger models.

In future work, we plan to: (i) study the impact of various hyperparameters (\eg learning rate) on the personalization process; and (ii) investigate new approaches to enhance personalization such as auto-regressive models, prompting, and reinforcement learning. Lastly, we also plan to investigate online approaches for regularly training these models on relevant and up-to-date code, and to cover further completion tasks (\eg modifying a single token instead of entire code lines or blocks). All code and data used in our study is publicly available~\cite{replication}.

	\section*{Acknowledgments}
	We are deeply grateful to the \href{https://i3us.us.es/}{I3US Institute} and the \href{https://score.us.es/}{SCORE Lab} at the University of Seville for providing us access to their HPC cluster, on which the experiments were run, and without which this research would not be possible. We acknowledge the financial support of the Swiss National Science Foundation for the PARSED project (SNF Project No.~219294).
	\clearpage
	
	\bibliographystyle{ACM-Reference-Format}
	\bibliography{main}

\end{document}